\documentclass[aps,prl,reprint,superscriptaddress,twocolumn,showkeys,amsmath,amssymb,longbibliography,floatfix,nobibnotes]{revtex4-2}

\usepackage{longtable}
\usepackage{morefloats}
\usepackage[dvips]{graphicx}
\usepackage{color}
\usepackage{epsfig,graphicx,amsfonts,amsbsy}
\usepackage{amsmath,amsfonts,amsthm,amssymb}
\usepackage{appendix}

\usepackage{bbm}
\usepackage{makeidx}
\usepackage{bm}
\usepackage{url}
\usepackage{verbatim}
\usepackage{braket}
\usepackage[bookmarksnumbered,pdfpagelabels=true,plainpages=false,colorlinks=true,linkcolor=black,citecolor=black,urlcolor=black]{hyperref}
\usepackage[rightcaption]{sidecap}
\usepackage{array}
\usepackage{booktabs}
\usepackage{multirow}
\usepackage{bbm}
\usepackage{tabularx}
\usepackage{cancel,soul,ulem}

\newcommand{\bs}[1]{\boldsymbol{#1}}

\graphicspath{{figures/}}

\makeatletter
\begin{document}
\title{Chiral magnon-polaron edge states in Heisenberg-Kitaev magnets}

\author{Jose D. Mella}
\affiliation{School of Engineering and Sciences, Universidad Adolfo Ib\'a\~nez, Santiago, Chile}
\author{L. E. F. Foa Torres}
\affiliation{Departamento de F\'isica, FCFM, Universidad de Chile, Santiago, Chile.}
\author{Roberto E. Troncoso}
\affiliation{School of Engineering and Sciences, Universidad Adolfo Ib\'a\~nez, Santiago, Chile}

\date{\today}

\begin{abstract}
The interplay of spin and lattice fluctuations in two-dimensional magnets without inversion symmetry is investigated. We find a general form for the magnetoelastic coupling between magnons and existing chiral phonons based on the symmetries of the crystalline lattice. We show that in hexagonal lattices, the coupling of magnons and chiral phonons derives from an anisotropic exchange spin model containing topological phases of magnons. Using the Heisenberg-Kitaev-$\Gamma$ model, we show how  magnon-polaron edge states with circular polarization arise from this interaction. Our findings exploit the polarization degrees of freedom in spin-lattice systems, thus setting the ground for the transfer of angular momentum between chiral phonons and magnons.
\begin{description}

\item[Keywords]
Magnon-phonon coupling, magnon-chiral phonon edge states, two dimensional materials, chiral phonons
\end{description}
\end{abstract}

\maketitle

\textit{Introduction.--} Chirality, an inherent handedness, emerges as a physical property that differ from its own mirror reflection. This concept is widespread in condensed matter physics, ranging from topological insulators \cite{kane2005quantum} and chiral anomaly in Weyl semimetals \cite{RylandsPRL2021}, to topological spin textures in magnetism \cite{Nagaosa2013,Gbel2021}. The absence of inversion symmetry in ordered crystals paves the way for the rising of chirality in more common systems as, lattice phononic vibrations  \cite{ChenIOP2018}. Chiral phonons describe circularly polarized vibrations, associated to rotations of atoms around their equilibrium positions, that carry pseudo-angular momentum \cite{ZhangPRL2014,ZhangPRL2015}. Lattice fluctuations with definitive handedness have been theoretically predicted in materials that harbor magnetism, via time-reversal-symmetry breaking \cite{ZhangPRL2014}, and in inversion-symmetry-broken systems such as, honeycomb \cite{ZhangPRL2014,ZhangPRL2015,XuPRB2018}, Kekulé \cite{LiuPRL2017}, kagome \cite{ChenPRB2019}, and triangular lattices \cite{XuJPCM2018}. Subsequent experimental confirmation \cite{Zhu2018} boosted significant attention, e.g., by their entanglement with photons \cite{chen2019entanglement}, interactions with electrons \cite{mella2023entangled}, large phonon magnetic moment induced by the interaction with spin-orbit exciton \cite{lujan2024spin}, interplay with electronic topology \cite{hernandez2023observation} and roles in superconductivity mediation \cite{gao2023chiral}. Recently, it has been demonstrated that these chiral modes can be manipulated by strong magnetic field \cite{BaydinPRL2022}, opening new avenues for their exploration and control.

Magnons, the quanta of spin fluctuations, mediate pure spin transport in magnetic insulators, where the low-dissipative and wave-like nature are exploited for future information communications technologies \cite{Chumak2015}. Besides potential applications, magnons and their bosonic roots allow for intriguing phenomena such as room temperature Bose–Einstein condensation \cite{Demokritov2006,bozhko2016supercurrent,Troncoso2012DynamicsAS}, quantum computing \cite{Chumak2015,Yuan2022}, topological matter \cite{Li2021,Zhuo2023}, and coherent hybridization with collective excitations such as, photons \cite{HueblPRL2013,ZhangPRL2014,TabuchiPRL2014}, plasmons \cite{GhoshPRB2023,DyrdalPRB2023,Costa2023}, and phonons \cite{Bozhko2020,Wang2023,Zheng2023}. Bridging spin and lattice fluctuations, phonons play a crucial role in the transport of spin angular momentum in magnetic insulators \cite{Bozhko2020}. The relaxation process of magnetization dynamics within a lattice is mainly mediated by the conversion between magnons and phonons, which is accompanied by the transfer of angular momentum between the spin and lattice system \cite{CornelissenPRB2016,NakanePRB2018,StreibPRL2018,StreibPRB2019,RuckriegelPRL2020,RuckriegelPRB2020,Bozhko2020,TroncosoPRB2020}. In intrinsically thermal phenomena, e.g., the spin Seebeck and heat conveyer effects \cite{bezuglyj2019spin,shklovskij2018role,an2013unidirectional}, it has been demonstrated that magnon-phonon coupling generates substantial Berry curvature that contributes to the thermal Hall effect \cite{sheikhi2021hybrid,park2020thermal,zhang2019thermal,li2022thermal}. Magnons, as well as phonons, are sensitive to the symmetries of the crystalline lattice of magnets \cite{Brinkman1967}. Although chiral effects have been recently investigated in 2D magnets \cite{Cui2023,ma2023chiral,WangPRB2023,yao2024conversion,delugas2023magnon}, a thorough understanding of systems that sustain the coupling of chiral phonons and magnons, and how chirality might be induced, is still under development.

In this Letter, we propose a general spin-lattice Hamiltonian sustaining coupling of magnons and chiral phonons in the absence of inversion symmetry, allowed by crystalline symmetries. Focusing in hexagonal lattices, we consider the Heisenberg-Kitaev-Gamma (HK$\Gamma$) model alongside the first nearest neighbor phonon dispersion. We examine the bulk and edge band structures in detail including magnon-phonon coupling, highlighting hybridization at crossing points, gaps, and the resulting edge states. Following the exploration of magnon-chiral phonon coupling. We identify how magnon-phonon coupling induces distinct magnon-polaron edge states. One set is characterized by the initial topological magnon edge states coupled with the circular polarization originating from the bulk-like phonon background. Simultaneously, we observe edge states that emerge solely due to magnon-phonon coupling, which are distinctly localized at the edges and exhibit a specific phonon polarization.

\textit{Hamiltonian model.--} We consider a two-dimensional ferromagnetic system with localized spins on a periodic lattice. The total Hamiltonian is ${\cal H}={\cal H}_S+{\cal H}_{ME}+{\cal H}_E$, with ${\cal H}_S$ and ${\cal H}_E$ the spin and elastic Hamiltonian, respectively. The magnetoelastic energy ${\cal H}_{ME}$ that couples spins and elastic deformations, originates from the lowest order expansion of the spin model defined on the lattice. The most general coupling between chiral-phonons and magnonic fluctuations can be determined phenomenologically from the magnetoelastic interaction, based on the symmetries of the crystal. To lowest order the microscopic Hamiltonian is $\mathcal{H}_{ME} =\sum_{\bs r \neq \bs r'} \sum_{\alpha\beta\gamma\lambda} B_{\bs r\bs r'}^{\alpha\beta\gamma\lambda} S_{\bs r}^{\alpha} S_{\bs r'}^{\beta} R_{\bs r\bs r'}^{\gamma\lambda}$, where ${\bs S}_{\bs r}$ is the spin at lattice site $\bs r$ and the discrete strain tensor is $R^{\gamma\lambda}_{\bs r\bs r'} = \left[\left( { r}^{\gamma} - {r'}^{\gamma} \right)\left( u_{\bs r}^{\lambda} - u_{\bs r'}^{\lambda} \right)+({\gamma \leftrightarrow\lambda})\right]/2|{\bs r} - \bs r' |^2$, where ${\bs u}_{\bs r}$ is the local displacement of an atom from its equilibrium position. The tensor $B_{\bs r\bs r'}^{\alpha\beta\gamma\lambda}$ is the magnetoelastic coupling, where the number of independent elements can be reduced with symmetry arguments
\footnote{The coefficients must be invariant under exchange of lattice site indices $i \leftrightarrow j$, so that there is only one tensor of coefficients per lattice pair $\{i,j\}$. Also, to be invariant under the transformations $\alpha \leftrightarrow \beta$ and $\gamma \leftrightarrow \lambda$. These symmetries reduce the number of coefficients to $36$ per lattice pair $\{i,j\}$.}. Based on the {\it Neumann's principle} \cite{birss1964symmetry}, the magnetoelastic Hamiltonian of a system with crystallographic point group $\mathcal{G}$ must be invariant under the symmetry operations $\mathcal{R} \in \mathcal{G}$, i.e. $\mathcal{H}_{ ME} [{\bs S}',{R}'] = \mathcal{H}_{ ME}[{\bs S},{R}]$, implying the condition ${\bs B}'_{\bs r\bs r'}={\mathcal{\bs R}}^T{\mathcal{\bs R}}^T{\bs B}_{\bs r\bs r'}{\mathcal{\bs R}}{\mathcal{\bs R}}$ for the magnetoelastic tensor. Specifying the crystalline system, we find the corresponding magnetoelastic coupling allowed by the symmetries, from which a general form for the coupling between chiral-phononic oscillations and magnons is determined. 

We next consider a hexagonal lattice in absence of inversion symmetry, e.g., by adding a staggering on the masses of atoms within phonon system, where the appearance of chiral-phononic oscillations is guaranteed \cite{ZhangPRL2015}. The symmetry group is $\mathcal{G}=D_{3h}$, and the resultant magnetoelastic Hamiltonian is $\mathcal{H}_{ ME}=\sum_{\langle \bs r\bs r' \rangle}{\cal J}^{\alpha\beta}_{\bs r\bs r'}S^\alpha_{\bs r}S^\beta_{\bs r'}$,
where ${\cal J}^{xx}_{\bs r\bs r'}={\tilde J}_{\bs r\bs r'}-{\tilde\Gamma}^{yy}_{\bs r\bs r'}$, ${\cal J}^{yy}_{\bs r\bs r'}={\tilde J}_{\bs r\bs r'}-{\tilde\Gamma}^{xx}_{\bs r\bs r'}$, and ${\cal J}^{zz}_{\bs r\bs r'}={\tilde J}_{\bs r\bs r'}+{\tilde K}_{\bs r\bs r'}$, where ${\tilde J}_{\bs r\bs r'}={\tilde J}\left(R^{xx}_{\bs r\bs r'}+R^{yy}_{\bs r\bs r'}\right)$, ${\tilde K}_{\bs r\bs r'}={\tilde K}^z\left(R^{xx}_{\bs r\bs r'}+R^{yy}_{\bs r\bs r'}\right)$, and ${\tilde\Gamma}^{\alpha\beta}_{\bs r\bs r'}={\tilde \Gamma}R^{\alpha\beta}_{\bs r\bs r'}$. We note that chiral phonons modes correspond to in-plane oscillations and thus, all terms proportional to $R^{z\lambda}_{\bs r\bs r'}(R^{\gamma z}_{\bs r\bs r'})$ can be disregarded, see SM for details. It is worth to stress that although the phenomenological derivation of $\mathcal{H}_{\rm ME}$ is ruled by the symmetries of the crystal, the couplings ${\tilde J}$, ${\tilde \Gamma}$ and $\tilde{K}^z$, originate microscopically from an anisotropic exchange spin model. This motivates us to consider the HK$\Gamma$ Hamiltonian \cite{JoshiPRB2018,McClartyPRB2018} as the reference spin model,
\begin{align}\label{eq:HK-model}
\mathcal{H}_{S}=\nonumber&J\sum_{\langle \bs r\bs r'\rangle}{\bs S_{\bs r}}\cdot {\bs S_{\bs r'}}-\sum_{\bs r}
{\bs h}\cdot{\bs S}_{\bs r}+\sum_{\langle {\bs r}{\bs r'}\rangle_\gamma}\left[2K^\gamma S_{\bs r}^\gamma S_{\bs r'}^\gamma\right.\\
&\qquad\left.+\Gamma^\gamma_{\alpha\beta}(S_{\bs r}^\alpha S_{\bs r'}^\beta+S_{\bs r}^\beta S_{\bs r'}^\alpha)\right],
\end{align} 
with $J$ the exchange coupling, $K^{\gamma}$ and $\Gamma^{\gamma}_{\alpha\beta}$ the bond-dependent coupling Kitaev and spin-anisotropic interaction, respectively. We parameterize $J=J_0\cos\phi$ and $K^{\gamma}=K=J_0\sin\phi$, where $J_0$ is a normalization exchange parameter and $\phi$ a parameter that set the ground state  \cite{JoshiPRB2018}. 
A magnetic field ${\bs h}$ saturates the magnetic state that is oriented an angle $\theta$ from $z$ direction. Although HK$\Gamma$-model exhibits a complex magnetic phase diagram, here we focus in a collinear ferromagnetic phase.

We now explore the low-energy spin and lattice fluctuations of the total Hamiltonian. Deviations around the ferromagnetic ground state are determined using Holstein-Primakoff (HP) bosons \cite{HolsteinPR1940}, through the spin-boson mapping ${S}^{+}_{\bs r}=\left(2S-a^{\dagger}_{\bs r}a_{\bs r}\right)^{1/2}a_{\bs r}$, ${S}^{-}_{\bs r}=\left(2S-a^{\dagger}_{\bs r}a_{\bs r}\right)^{1/2}a^{\dagger}_{\bs r}$ and ${S}^{z}_{\bs r}=S-a^{\dagger}_{\bs r}a_{\bs r}$. Thus, expanding the spin operators as a series in $1/S$ \footnote{It is convenient to choose the quantization axis along the saturation magnetization, such that ${\bs S}_{\bs r}={\cal R}_y(\theta)\tilde{\bs S}_{\bs r}$, with ${\cal R}_y(\theta)$ the rotation matrix in an angle $\theta$ around $y$ axis $S^x=\tilde{S}^{x'}\cos(\theta)+\tilde{S}^{z'}\sin(\theta)$, $S^y=\tilde{S}^{y'}$ and $S^z=-\tilde{S}^{x'}\sin(\theta)+\tilde{S}^{z'}\cos(\theta)$}, the Hamiltonian for non-interacting magnons in Fourier space is ${\cal H}_m=\sum_{\bs k}\Psi_{\bs k}^\dagger H^m_{\bs k}\Psi_{\bs k}$, with the matrix
\begin{align}
{H}^m_{\bs k}=\frac{S}{2}\left(\begin{array}{cc}
A_{\bs k}  & B_{\bs k} \\
B^{\dagger}_{\bs k}  & A_{\bs k}
\end{array}\right),
\end{align}
and the field operator $\Psi_{\bs k}=(a_{\bs k},b_{\bs k},a_{-\bs k}^\dagger,b_{-\bs k}^\dagger)^T$, with the operator $a_{\bs k}$ and $b_{\bs k}$ defined on the ${\cal A}$- and ${\cal B}$-sublattice, respectively. The matrix elements $A^{ij}_{\bs k}=\kappa_0\delta_{ij}+\kappa_{1,(-1)^j\bs k}\delta^{a}_{ij}$ and $B^{ij}_{\bs k}=\kappa_{2,(-1)^j\bs k}\delta^{a}_{ij}$, being $\delta^{a}_{ij}$ the $2\times 2$ anti-diagonal matrix, and the coefficients

\begin{align*}
\kappa_0=&-3J-2K+\frac{h_x}{S}\sin\theta+\frac{h_z}{S} \cos\theta, \\
\kappa_{1\bs k}=&J\gamma_{\bs k}+K(\cos^2\theta+\sin^2\theta e^{i{\bs k} \cdot {\bs a_1}}+e^{i{\bs k} \cdot {\bs a_2}}),\\
\kappa_{2\bs k}=&K(\cos^2\theta+\sin^2\theta e^{i{\bs k} \cdot {\bs a_1}}-e^{i {\bs k} \cdot {\bs a_2}}) +i\Gamma_z \cos\theta e^{i{\bs k} \cdot {\bs a_1}},
\end{align*}
where $\gamma_{\bs k}=1+\sum_{n}e^{i{\bs k}\cdot{\bs a}_n}$ with $n\in\{1,2\}$. For simplicity, we focus on the case where $\Gamma_x=\Gamma_y=0$ (see SM for the full Hamiltonian). Note that when the magnetic state is polarized along the $[0,0,1]$-direction, i.e., $\theta=0$, the coefficients reduce to the Ref. \cite{JoshiPRB2018}. Lattice vibrations are described by the Hamiltonian for phonons $\mathcal{H}_{ph}=\sum_{\bs k,\xi} \hbar\omega_{\bs k\xi}c_{\bs q\xi}^\dagger c_{\bs q\xi}$, where the frequencies are obtained from the standard eigenvalue problem $\bm{D}(\bs k)\bm{\bar{u}}_{\bs k}=\omega_{\bs k}^2\bm{\bar{u}}_{\bs k}$, see SM for details. Assuming that inversion symmetry is broken, a non-zero pseudoangular momentum is induced in the vicinity of the valleys, which is linked to the circular motion exhibited by the sublattices \cite{ZhangPRL2015}. We consider the next basis, $\ket{{R}_A}=(1,i,0,0)^T/\sqrt{2}$, $\ket{L_A}=(1,-i,0,0)^T/\sqrt{2}$, $\ket{R_B}=(0,0,1,i)^T/\sqrt{2}$ and $\ket{L_B}=(0,0,1,-i)^T/\sqrt{2}$, that are associated with the right- and left-handed circular motion of sublattices ${\cal A}$ and ${\cal B}$. The phonon polarization vector can be written as $\bs{\mu}=\sum_{\alpha}\varepsilon_{R_\alpha}\ket{R_\alpha}+\varepsilon_{L_\alpha}\ket{L_\alpha}$, where the circular polarization of the phonon mode is $s^z_{\text{ph}}=\hbar l_{\text{ph}}$, with $l_{\text{ph}}=\sum_{\alpha}(|\varepsilon_{R_\alpha}|^2-|\varepsilon_{L_\alpha}|^2)$.
\begin{figure*}
    \centering
\includegraphics[width=\textwidth]{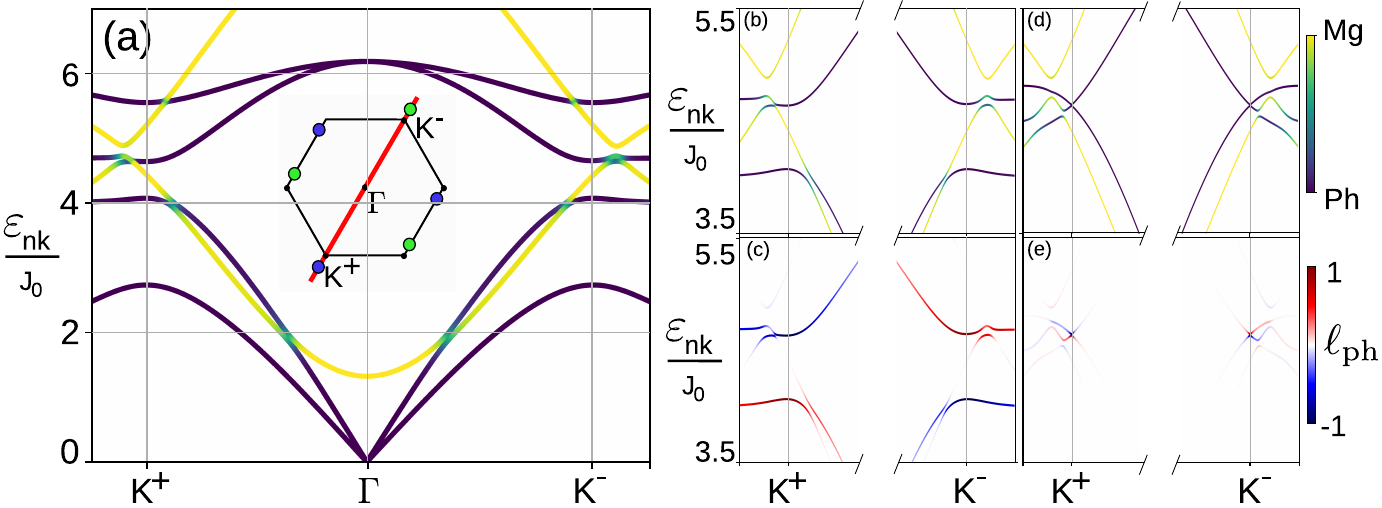}
    \caption{Representative magnon-phonon dispersion with (d,e) and without (a,b,c) inversion symmetry breaking within the phonon system. Panel (a) shows the dispersion projected in magnons and phonons, and the inset represents the $k$-path followed, where the blue and green dots indicate the magnon band gap positions. The zoom around the $K^{\pm}$ projected in boson type (b,d) and phonon polarization (c,e) shows the differences in the band gaps due to the inversion symmetry breaking. The parameters used in all figures are $a_0=1$, $S=1$, $\phi=5\pi/4$, $\Gamma_z=0.7J_0$, $h_x=h_z=J_0S$, $K_0=\Gamma_0=0.075$, $\phi_L=11.5J_0$ and $\phi_T=2.875J_0$. We break the inversion symmetry in the phonon system setting $m_2=1.1m_1$, while $m_1=m_2$ when inversion is preserved.}
    \label{fig:bulks-mph}
\end{figure*}

The coupling between magnons and phonons derives from the magnetoelastic Hamiltonian found for the lattice with $D_{3h}$ point group of symmetries. The local displacement, ${\bs u}_{\bs r}$, is expressed in terms of the phonon operators \footnote{${\bs u}_{\bs r}=\sum_{\bs k,\xi} {\bs \epsilon}_{\bs k\xi}\sqrt{{\hbar}/{2m_{\bs r}\omega_{\bs k\xi}N}}\left(c_{\bs k\xi}^\dagger+c_{-\bs k\xi}\right)e^{i\bs k\cdot {\bs R}_{\bs r}}$} to find the strain tensor $R_{{\bs r}{\bs r'}}^{\gamma \lambda}= \sum_{\bs k,\xi} \sqrt{\frac{\hbar}{2N}}\left(c_{\bs k\xi}^\dagger+c_{-\bs k\xi}\right)e^{i\bs k\cdot {\bs R}_{{\bs r}}}L^{\gamma \lambda}_{{\bs \delta}{\bs k}\xi}$, with $m_{\bs r}$ the local mass, $N$ the number of lattice sites, $\bs R_{\bs r}$ the unit cell vector, and $\bs \delta={\bs r'}-{\bs r}$ the relative position. The tensor $L^{\gamma \lambda}_{{\bs \delta}{\bs k}\xi}=(\tilde{\mu}^{\gamma \lambda}_{\bs k \xi}-\tilde{ \nu}^{\gamma \lambda}_{\bs k \xi}e^{i\bs k \cdot \Delta{\bs R}_{{\bs r}{\bs r'}}})/2\delta^2\sqrt{\omega_{\bs k\xi}}$, where $\tilde{\mu}^{\gamma \lambda}_{\bs k \xi}=({\delta}^\lambda{\mu_{\bs k \xi}^\gamma+{\delta}^\gamma{ \mu_{\bs k \xi}^\lambda}})/\sqrt{m_{\bs r}}$ and $\tilde{ \nu}^{\gamma \lambda}_{\bs k \xi}=({\delta}^\lambda{ \nu_{\bs k \xi}^\gamma+{\delta}^\gamma{ \nu_{\bs k \xi}^\lambda}})/\sqrt{m_{\bs r'}}$, and $\Delta {\bs R}_{{\bs r}{\bs r'}}=\bs R_{\bs r'}-\bs R_{\bs r}$. Here, $\bs{\mu}$ and $\bs{\nu}$ correspond to the polarization vector for sublattices ${\cal A}$ and ${\cal B}$, respectively. Thus, choosing the quantization axis along the saturation magnetization and expanding $\mathcal{H}_{ME}$ up to second order in bosonic operators (complete spin-phonon Hamiltonian is detailed at SM), we employ the HP representation to derive the magnon-phonon Hamiltonian, obtaining
\begin{align}
\mathcal{H}_{mp}\nonumber=\sum_{\bs k \xi}&\left(N^*_{-\bs k \xi}a_{\bs k}^\dagger+N_{\bs k \xi}a_{-\bs k}+M_{-\bs k \xi}^* b_{\bs k}^\dagger+M_{\bs k \xi}b_{-\bs k}\right)\\
&\qquad\qquad\qquad\qquad\qquad \times\left(c_{-\bs k\xi}^\dagger+c_{\bs k\xi}\right),
\end{align}
where the $\bs k$-dependent magnon-phonon couplings are
\begin{align*}
N_{\bs k \xi \theta}&=N^a_{\bs k\xi}\sin(2\theta)+iN^b_{\bs k\xi}\sin(\theta)+N^c_{\bs k\xi}\sin(2\theta),\\
M_{\bs k \xi}&=e^{i\bs k\cdot \bs a_1}\left[N^a_{\bs k\xi}\sin(2\theta)+iN^b_{\bs k\xi}\sin(\theta)\right]+N^c_{\bs k\xi}\sin(2\theta),
\end{align*}
with $N^a_{\bs k\xi}=-K_0\left(\tilde{L}^{xx}_{\bs \delta \bs k \xi}+\tilde{L}^{yy}_{\bs \delta \bs k \xi}\right)$, $N^b_{\bs k\xi }=-\Gamma_0\tilde{L}^{xy}_{\bs \delta \bs k \xi}$, $N^c_{\bs k\xi}=-\Gamma_0L^{yy}_{\bs \delta \bs k \xi}$, $\tilde{L}_{\bs \delta \bs k \xi}=\tilde{L}_{\bs \delta -\bs k \xi}$, and the coefficients $K_0=\tilde{K}^z\sqrt{S^3\hbar}/2$ and $\Gamma_0=\tilde{\Gamma}_{\bs r \bs r'}\sqrt{S^3\hbar}/2$. We note that when the quantization axis is perpendicular to the plane ($\theta = 0$), the magnon-phonon coupling vanishes. Consequently, there is no coupling between in-plane phonon modes and magnons. To enable the coupling between magnons and chiral phonons (in-plane modes), we consider $\theta \neq 0$. The total Hamiltonian for the system of magnons and phonons reads ${\cal H}_T={\cal H}_m+{\cal H}_{mp}+{\cal H}_p$ that might be written as ${\cal H}_T=\sum_{\bs k}\Phi^{\dagger}_{\bs k}H_{\bs k}\Phi_{\bs k}$ and $\Phi_{\bs k}=\left(\Psi^m_{\bs k},\Phi^p_{\bs k}\right)^T$, where $\Psi^{m}_{\bs k}$ and $\Phi^p_{\bs k}$ are the magnonic and phononic operators respectively. The bosonic Hamiltonian is diagonalized by the Bogoliubov transformation $\left(\Psi^m_{\bs k},\Phi^p_{\bs k}\right)^T={\mathcal{T}}_{\bs k}({\bs \alpha}_{\bs k},{\bs \alpha}^{\dagger}_{-\bs k})^T$ using the Colpa method (for details see SM) \cite{colpa1978diagonalization}, with ${\mathcal{T}}$ the paraunitary transformation that satisfy ${\mathcal{T}}^{\dagger}\zeta{\mathcal{T}}=\zeta$ to guarantee the bosonic commutation relation $\left[{\bs \alpha},{\bs \alpha}^{\dagger}\right]=\mathbb{I}\otimes\sigma_z=\zeta$ \cite{colpa1978diagonalization}. The diagonalized magnon-phonon Hamiltonian is written as ${\cal H}_T=\sum_{n,\bs k}{\cal E}_{n\bs k}\alpha^{\dagger}_{n\bs k}\alpha_{n\bs k}$ with ${\epsilon}_{n\bs k}$ the energy for the $n^{\text{th}}$-band. 

\begin{figure*}
    \centering
\includegraphics[width=\textwidth]{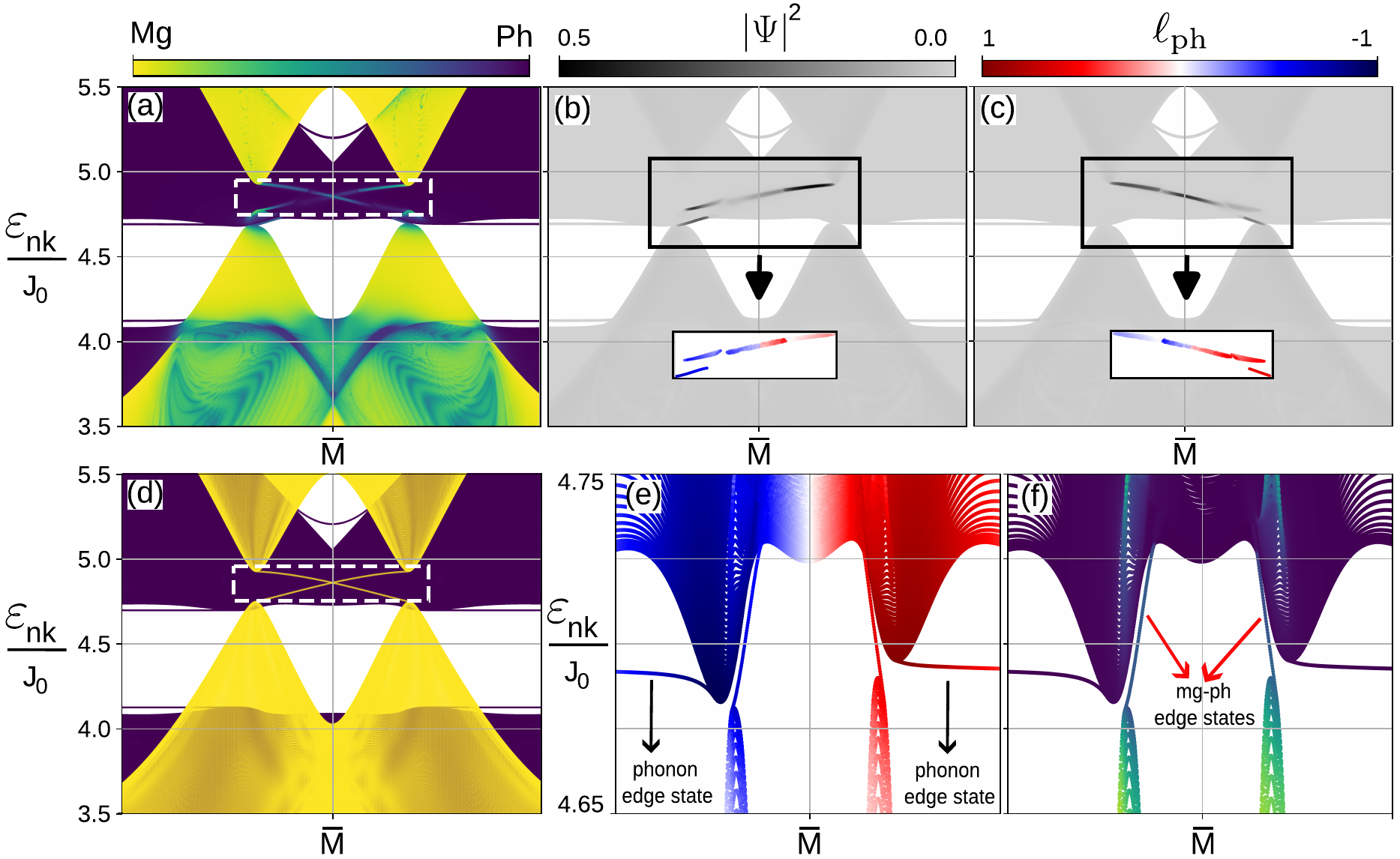}
    \caption{Zig-zag nanoribbon dispersion of 300 unit-cell for the magnon-phonon system. Panel (a,b) shows the bosons contribution with (a) and without (b) coupling. The light-purple energy zone in panel (d) shows the topological magnon edge states. (b,c) depict the bottom and top edge projection adding a magnon projection, where the subset is the phonon polarization of these states. Panel (e,d) shows the magnon-polaron edge states originated from the magnon-phonon crossings. The yellow-green scale shows the magnon and phonon contribution to the state, while grey-scale shows the wave function projected into edges and magnon-part (5 unit cells). The red-blue scale depicted the phonon polarization $l_{ph}$.}
    \label{fig:mgph_edge}
\end{figure*}

\textit{Magnon-phonon hybridization.--} We begin our discussion by examining the effects of magnon-phonon coupling when inversion symmetry is preserved within the phonon system, i.e., $m_1 = m_2$. In general, this interaction induces the formation of band gaps at all band crossings between the phonon and magnon bands \cite{sheikhi2021hybrid,Cui2023}. A representative magnon-phonon dispersion and the emergence of different gaps at several $k$-points in the Brillouin zone is shown at Fig. \ref{fig:bulks-mph}(a) for $\theta=\pi/4$. The magnon(phonon)-like dispersions are seen by the yellow(purple) line. The Kitaev term ($K$) and $\theta$ moves the non-equivalent Dirac points out of high-symmetry point (depicted in green and blue dot in the subset of Fig.~\ref{fig:mgph_edge}(a), while $\Gamma^z$ introduces a topological band-gap \cite{JoshiPRB2018}. Notably, the gap openings near the nonequivalent ${\text K}^+$ and ${\text K}^-$ points exhibit similar magnitudes and present mirrored behaviors. The magnon-polaron states near the crossing also present a definite circular polarization and it is opposite at each Dirac point, see Fig. \ref{fig:bulks-mph}(e). This effect is related to the angular momentum transfer between magnons and phonons \cite{RuckriegelPRL2020}.

To understand the effect of the phonon polarization in the magnon-phonon coupling, we explore the system when the inversion symmetry is broken for the phonons, i.e., $m_1 \leq m_2$. The phonon modes at the Dirac points exhibit a definite chirality, as seen in Fig.~\ref{fig:bulks-mph}(b). The dependence of the phonon mode on the magnon-phonon coupling suggests differences in its magnitude associated with the chiral modes, which may influence the behavior of the band gap. We focus on a detailed view of the magnon-phonon band gap shown in Fig.~\ref{fig:bulks-mph}(b,c). We observe that the mirror property of the band gap discussed previously is no longer present, and differences in the band gap arise. This discrepancy can be attributed to the inversion symmetry breaking in the phonon system, which extrapolates the symmetry breaking to the coupling through the phonon modes. This generates different amplitude couplings in the different inequivalent valleys ($K^{\pm}$) and in consequence, leads to a different band gap. This results enlighten the critical role of phonon polarization in modulating valley-dependent magnon properties in the presence of magnon-phonon coupling.

{\textit{Chiral magnon-polaron edge-states.--} We now investigate the hybrid states localized at the edges of a nanoribbon with a zig-zag edge termination along the $x$-direction. The resulting spectrum, displayed in Fig. \ref{fig:mgph_edge}(a), reveals the emergence of two sets of magnon-polaron edge-modes. In the absence of coupling, we identify the topological magnon edge-states (white box) and the phonon background, consistent with previous calculations \cite{JoshiPRB2018,savin2010vibrational}, as shown in Fig. \ref{fig:mgph_edge}(d). The interaction between topological magnon edge states and the phonon background leads to the formation of magnon-polaron states, observed in the white box of Fig. \ref{fig:mgph_edge}(a). These modes consist of a topological magnon and a bulk-like phonon, each polarized up or down depending on the energy and specific ${\bs k}$-point (Figs. \ref{fig:mgph_edge}(b,c)). While the magnon component remains localized at the edge, the phonon part acquires polarization from the bulk-like phonon background, which acts as a dissipative reservoir \cite{broadening2020}. Consequently, these states exhibit broadening and reduced robustness of the topological magnon part due to the bulk-like phonon component allowing backscattering with the background, e.g., in the presence of disorder. At a given energy, these magnon-polaron edge-states have definite circular polarization, with the opposite edge exhibiting a state with opposite direction of propagation and polarization. Another set of hybrid edge-states, resulting from the magnon-phonon coupling, is depicted in Fig. \ref{fig:mgph_edge}(e,f) and highlighted by red arrows. These states also exhibit definite circular polarization and are localized at different edges, as shown in Fig. \ref{fig:mgph_edge}(b,c).}

In summary, our study reveals the crystal symmetries-based spin-phonon Hamiltonian that support coupling between chiral phonons and magnons. Within the HK$\Gamma$ model, we show the effects from phonon chirality in the bulk and edge band structure. We observe the formation of two sets of magnon-polaron edge states, each of one with a definite circular phonon polarization. These findings pave the way for exploiting phonon chirality, potentially enabling enhanced control over entanglement and state manipulation in solid-state systems.

\section{ACKNOWLEDGMENTS}
R.E.T and J.M. thanks funding from Fondecyt Regular 1230747. L.F.T. acknowledges financial support from ANID Fondecyt Regular 1211038, the EU Horizon 2020 research and innovation program under the Marie-Sklodowska-Curie Grant Agreement No. 873028 (HYDROTRONICS Project), and the support of The Abdus Salam International Centre for Theoretical Physics and the Simons Foundation.

\section{AUTHOR CONTRIBUTION}
{JM and RET discussed the initial idea for the project. JM carried all the numeric calculations, while RET independently verified these calculations with his own results. All authors discussed the results and wrote the manuscript.}

\section{REFERENCES}
\bibliography{magnonphonon}

\begin{thebibliography}{63}%
\makeatletter
\providecommand \@ifxundefined [1]{%
 \@ifx{#1\undefined}
}%
\providecommand \@ifnum [1]{%
 \ifnum #1\expandafter \@firstoftwo
 \else \expandafter \@secondoftwo
 \fi
}%
\providecommand \@ifx [1]{%
 \ifx #1\expandafter \@firstoftwo
 \else \expandafter \@secondoftwo
 \fi
}%
\providecommand \natexlab [1]{#1}%
\providecommand \enquote  [1]{``#1''}%
\providecommand \bibnamefont  [1]{#1}%
\providecommand \bibfnamefont [1]{#1}%
\providecommand \citenamefont [1]{#1}%
\providecommand \href@noop [0]{\@secondoftwo}%
\providecommand \href [0]{\begingroup \@sanitize@url \@href}%
\providecommand \@href[1]{\@@startlink{#1}\@@href}%
\providecommand \@@href[1]{\endgroup#1\@@endlink}%
\providecommand \@sanitize@url [0]{\catcode `\\12\catcode `\$12\catcode
  `\&12\catcode `\#12\catcode `\^12\catcode `\_12\catcode `\%12\relax}%
\providecommand \@@startlink[1]{}%
\providecommand \@@endlink[0]{}%
\providecommand \url  [0]{\begingroup\@sanitize@url \@url }%
\providecommand \@url [1]{\endgroup\@href {#1}{\urlprefix }}%
\providecommand \urlprefix  [0]{URL }%
\providecommand \Eprint [0]{\href }%
\providecommand \doibase [0]{http://dx.doi.org/}%
\providecommand \selectlanguage [0]{\@gobble}%
\providecommand \bibinfo  [0]{\@secondoftwo}%
\providecommand \bibfield  [0]{\@secondoftwo}%
\providecommand \translation [1]{[#1]}%
\providecommand \BibitemOpen [0]{}%
\providecommand \bibitemStop [0]{}%
\providecommand \bibitemNoStop [0]{.\EOS\space}%
\providecommand \EOS [0]{\spacefactor3000\relax}%
\providecommand \BibitemShut  [1]{\csname bibitem#1\endcsname}%
\let\auto@bib@innerbib\@empty
\bibitem [{\citenamefont {Kane}\ and\ \citenamefont
  {Mele}(2005)}]{kane2005quantum}%
  \BibitemOpen
  \bibfield  {author} {\bibinfo {author} {\bibfnamefont {C.~L.}\ \bibnamefont
  {Kane}}\ and\ \bibinfo {author} {\bibfnamefont {E.~J.}\ \bibnamefont
  {Mele}},\ }\bibfield  {title} {\enquote {\bibinfo {title} {Quantum spin hall
  effect in graphene},}\ }\href {\doibase 10.1103/PhysRevLett.95.226801}
  {\bibfield  {journal} {\bibinfo  {journal} {Phys. Rev. Lett.}\ }\textbf
  {\bibinfo {volume} {95}},\ \bibinfo {pages} {226801} (\bibinfo {year}
  {2005})}\BibitemShut {NoStop}%
\bibitem [{\citenamefont {Rylands}\ \emph {et~al.}(2021)\citenamefont
  {Rylands}, \citenamefont {Parhizkar}, \citenamefont {Burkov},\ and\
  \citenamefont {Galitski}}]{RylandsPRL2021}%
  \BibitemOpen
  \bibfield  {author} {\bibinfo {author} {\bibfnamefont {C.}~\bibnamefont
  {Rylands}}, \bibinfo {author} {\bibfnamefont {A.}~\bibnamefont {Parhizkar}},
  \bibinfo {author} {\bibfnamefont {A.~A.}\ \bibnamefont {Burkov}}, \ and\
  \bibinfo {author} {\bibfnamefont {V.}~\bibnamefont {Galitski}},\ }\bibfield
  {title} {\enquote {\bibinfo {title} {Chiral anomaly in interacting condensed
  matter systems},}\ }\href {\doibase 10.1103/PhysRevLett.126.185303}
  {\bibfield  {journal} {\bibinfo  {journal} {Phys. Rev. Lett.}\ }\textbf
  {\bibinfo {volume} {126}},\ \bibinfo {pages} {185303} (\bibinfo {year}
  {2021})}\BibitemShut {NoStop}%
\bibitem [{\citenamefont {Nagaosa}\ and\ \citenamefont
  {Tokura}(2013)}]{Nagaosa2013}%
  \BibitemOpen
  \bibfield  {author} {\bibinfo {author} {\bibfnamefont {N.}~\bibnamefont
  {Nagaosa}}\ and\ \bibinfo {author} {\bibfnamefont {Y.}~\bibnamefont
  {Tokura}},\ }\bibfield  {title} {\enquote {\bibinfo {title} {Topological
  properties and dynamics of magnetic skyrmions},}\ }\href
  {http://dx.doi.org/10.1038/nnano.2013.243} {\bibfield  {journal} {\bibinfo
  {journal} {Nat. Nanotechnol.}\ }\textbf {\bibinfo {volume} {8}},\ \bibinfo
  {pages} {899–911} (\bibinfo {year} {2013})}\BibitemShut {NoStop}%
\bibitem [{\citenamefont {G\"{o}bel}\ \emph {et~al.}(2021)\citenamefont
  {G\"{o}bel}, \citenamefont {Mertig},\ and\ \citenamefont
  {Tretiakov}}]{Gbel2021}%
  \BibitemOpen
  \bibfield  {author} {\bibinfo {author} {\bibfnamefont {B.}~\bibnamefont
  {G\"{o}bel}}, \bibinfo {author} {\bibfnamefont {I.}~\bibnamefont {Mertig}}, \
  and\ \bibinfo {author} {\bibfnamefont {O.~A.}\ \bibnamefont {Tretiakov}},\
  }\bibfield  {title} {\enquote {\bibinfo {title} {Beyond skyrmions: Review and
  perspectives of alternative magnetic quasiparticles},}\ }\href {\doibase
  10.1016/j.physrep.2020.10.001} {\bibfield  {journal} {\bibinfo  {journal}
  {Phys. Rep.}\ }\textbf {\bibinfo {volume} {895}},\ \bibinfo {pages} {1–28}
  (\bibinfo {year} {2021})}\BibitemShut {NoStop}%
\bibitem [{\citenamefont {Chen}\ \emph {et~al.}(2018)\citenamefont {Chen},
  \citenamefont {Zhang}, \citenamefont {Niu},\ and\ \citenamefont
  {Zhang}}]{ChenIOP2018}%
  \BibitemOpen
  \bibfield  {author} {\bibinfo {author} {\bibfnamefont {H.}~\bibnamefont
  {Chen}}, \bibinfo {author} {\bibfnamefont {W.}~\bibnamefont {Zhang}},
  \bibinfo {author} {\bibfnamefont {Q.}~\bibnamefont {Niu}}, \ and\ \bibinfo
  {author} {\bibfnamefont {L.}~\bibnamefont {Zhang}},\ }\bibfield  {title}
  {\enquote {\bibinfo {title} {Chiral phonons in two-dimensional materials},}\
  }\href {\doibase 10.1088/2053-1583/aaf292} {\bibfield  {journal} {\bibinfo
  {journal} {2d Mater.}\ }\textbf {\bibinfo {volume} {6}},\ \bibinfo {pages}
  {012002} (\bibinfo {year} {2018})}\BibitemShut {NoStop}%
\bibitem [{\citenamefont {Zhang}\ and\ \citenamefont
  {Niu}(2014)}]{ZhangPRL2014}%
  \BibitemOpen
  \bibfield  {author} {\bibinfo {author} {\bibfnamefont {L.}~\bibnamefont
  {Zhang}}\ and\ \bibinfo {author} {\bibfnamefont {Q.}~\bibnamefont {Niu}},\
  }\bibfield  {title} {\enquote {\bibinfo {title} {Angular momentum of phonons
  and the einstein--de haas effect},}\ }\href {\doibase
  10.1103/PhysRevLett.112.085503} {\bibfield  {journal} {\bibinfo  {journal}
  {Phys. Rev. Lett.}\ }\textbf {\bibinfo {volume} {112}},\ \bibinfo {pages}
  {085503} (\bibinfo {year} {2014})}\BibitemShut {NoStop}%
\bibitem [{\citenamefont {Zhang}\ and\ \citenamefont
  {Niu}(2015)}]{ZhangPRL2015}%
  \BibitemOpen
  \bibfield  {author} {\bibinfo {author} {\bibfnamefont {L.}~\bibnamefont
  {Zhang}}\ and\ \bibinfo {author} {\bibfnamefont {Q.}~\bibnamefont {Niu}},\
  }\bibfield  {title} {\enquote {\bibinfo {title} {Chiral phonons at
  high-symmetry points in monolayer hexagonal lattices},}\ }\href {\doibase
  10.1103/PhysRevLett.115.115502} {\bibfield  {journal} {\bibinfo  {journal}
  {Phys. Rev. Lett.}\ }\textbf {\bibinfo {volume} {115}},\ \bibinfo {pages}
  {115502} (\bibinfo {year} {2015})}\BibitemShut {NoStop}%
\bibitem [{\citenamefont {Xu}\ \emph {et~al.}(2018{\natexlab{a}})\citenamefont
  {Xu}, \citenamefont {Chen},\ and\ \citenamefont {Zhang}}]{XuPRB2018}%
  \BibitemOpen
  \bibfield  {author} {\bibinfo {author} {\bibfnamefont {X.}~\bibnamefont
  {Xu}}, \bibinfo {author} {\bibfnamefont {H.}~\bibnamefont {Chen}}, \ and\
  \bibinfo {author} {\bibfnamefont {L.}~\bibnamefont {Zhang}},\ }\bibfield
  {title} {\enquote {\bibinfo {title} {Nondegenerate chiral phonons in the
  brillouin-zone center of $\sqrt{3}\ifmmode\times\else\texttimes\fi{}\sqrt{3}$
  honeycomb superlattices},}\ }\href {\doibase 10.1103/PhysRevB.98.134304}
  {\bibfield  {journal} {\bibinfo  {journal} {Phys. Rev. B}\ }\textbf {\bibinfo
  {volume} {98}},\ \bibinfo {pages} {134304} (\bibinfo {year}
  {2018}{\natexlab{a}})}\BibitemShut {NoStop}%
\bibitem [{\citenamefont {Liu}\ \emph {et~al.}(2017)\citenamefont {Liu},
  \citenamefont {Lian}, \citenamefont {Li}, \citenamefont {Xu},\ and\
  \citenamefont {Duan}}]{LiuPRL2017}%
  \BibitemOpen
  \bibfield  {author} {\bibinfo {author} {\bibfnamefont {Y.}~\bibnamefont
  {Liu}}, \bibinfo {author} {\bibfnamefont {C.-S.}\ \bibnamefont {Lian}},
  \bibinfo {author} {\bibfnamefont {Y.}~\bibnamefont {Li}}, \bibinfo {author}
  {\bibfnamefont {Y.}~\bibnamefont {Xu}}, \ and\ \bibinfo {author}
  {\bibfnamefont {W.}~\bibnamefont {Duan}},\ }\bibfield  {title} {\enquote
  {\bibinfo {title} {Pseudospins and topological effects of phonons in a
  kekul\'e lattice},}\ }\href {\doibase 10.1103/PhysRevLett.119.255901}
  {\bibfield  {journal} {\bibinfo  {journal} {Phys. Rev. Lett.}\ }\textbf
  {\bibinfo {volume} {119}},\ \bibinfo {pages} {255901} (\bibinfo {year}
  {2017})}\BibitemShut {NoStop}%
\bibitem [{\citenamefont {Chen}\ \emph
  {et~al.}(2019{\natexlab{a}})\citenamefont {Chen}, \citenamefont {Wu},
  \citenamefont {Yang}, \citenamefont {Li},\ and\ \citenamefont
  {Zhang}}]{ChenPRB2019}%
  \BibitemOpen
  \bibfield  {author} {\bibinfo {author} {\bibfnamefont {H.}~\bibnamefont
  {Chen}}, \bibinfo {author} {\bibfnamefont {W.}~\bibnamefont {Wu}}, \bibinfo
  {author} {\bibfnamefont {S.~A.}\ \bibnamefont {Yang}}, \bibinfo {author}
  {\bibfnamefont {X.}~\bibnamefont {Li}}, \ and\ \bibinfo {author}
  {\bibfnamefont {L.}~\bibnamefont {Zhang}},\ }\bibfield  {title} {\enquote
  {\bibinfo {title} {Chiral phonons in kagome lattices},}\ }\href {\doibase
  10.1103/PhysRevB.100.094303} {\bibfield  {journal} {\bibinfo  {journal}
  {Phys. Rev. B}\ }\textbf {\bibinfo {volume} {100}},\ \bibinfo {pages}
  {094303} (\bibinfo {year} {2019}{\natexlab{a}})}\BibitemShut {NoStop}%
\bibitem [{\citenamefont {Xu}\ \emph {et~al.}(2018{\natexlab{b}})\citenamefont
  {Xu}, \citenamefont {Zhang}, \citenamefont {Wang},\ and\ \citenamefont
  {Zhang}}]{XuJPCM2018}%
  \BibitemOpen
  \bibfield  {author} {\bibinfo {author} {\bibfnamefont {X.}~\bibnamefont
  {Xu}}, \bibinfo {author} {\bibfnamefont {W.}~\bibnamefont {Zhang}}, \bibinfo
  {author} {\bibfnamefont {J.}~\bibnamefont {Wang}}, \ and\ \bibinfo {author}
  {\bibfnamefont {L.}~\bibnamefont {Zhang}},\ }\bibfield  {title} {\enquote
  {\bibinfo {title} {Topological chiral phonons in center-stacked bilayer
  triangle lattices},}\ }\href {\doibase 10.1088/1361-648X/aabf5e} {\bibfield
  {journal} {\bibinfo  {journal} {J. Phys.: Condens. Matter.}\ }\textbf
  {\bibinfo {volume} {30}},\ \bibinfo {pages} {225401} (\bibinfo {year}
  {2018}{\natexlab{b}})}\BibitemShut {NoStop}%
\bibitem [{\citenamefont {Zhu}\ \emph {et~al.}(2018)\citenamefont {Zhu},
  \citenamefont {Yi}, \citenamefont {Li}, \citenamefont {Xiao}, \citenamefont
  {Zhang}, \citenamefont {Yang}, \citenamefont {Kaindl}, \citenamefont {Li},
  \citenamefont {Wang},\ and\ \citenamefont {Zhang}}]{Zhu2018}%
  \BibitemOpen
  \bibfield  {author} {\bibinfo {author} {\bibfnamefont {H.}~\bibnamefont
  {Zhu}}, \bibinfo {author} {\bibfnamefont {J.}~\bibnamefont {Yi}}, \bibinfo
  {author} {\bibfnamefont {M.-Y.}\ \bibnamefont {Li}}, \bibinfo {author}
  {\bibfnamefont {J.}~\bibnamefont {Xiao}}, \bibinfo {author} {\bibfnamefont
  {L.}~\bibnamefont {Zhang}}, \bibinfo {author} {\bibfnamefont {C.-W.}\
  \bibnamefont {Yang}}, \bibinfo {author} {\bibfnamefont {R.~A.}\ \bibnamefont
  {Kaindl}}, \bibinfo {author} {\bibfnamefont {L.-J.}\ \bibnamefont {Li}},
  \bibinfo {author} {\bibfnamefont {Y.}~\bibnamefont {Wang}}, \ and\ \bibinfo
  {author} {\bibfnamefont {X.}~\bibnamefont {Zhang}},\ }\bibfield  {title}
  {\enquote {\bibinfo {title} {Observation of chiral phonons},}\ }\href
  {\doibase 10.1126/science.aar2711} {\bibfield  {journal} {\bibinfo  {journal}
  {Science}\ }\textbf {\bibinfo {volume} {359}},\ \bibinfo {pages} {579–582}
  (\bibinfo {year} {2018})}\BibitemShut {NoStop}%
\bibitem [{\citenamefont {Chen}\ \emph
  {et~al.}(2019{\natexlab{b}})\citenamefont {Chen}, \citenamefont {Lu},
  \citenamefont {Dubey}, \citenamefont {Yao}, \citenamefont {Liu},
  \citenamefont {Wang}, \citenamefont {Xiong}, \citenamefont {Zhang},\ and\
  \citenamefont {Srivastava}}]{chen2019entanglement}%
  \BibitemOpen
  \bibfield  {author} {\bibinfo {author} {\bibfnamefont {X.}~\bibnamefont
  {Chen}}, \bibinfo {author} {\bibfnamefont {X.}~\bibnamefont {Lu}}, \bibinfo
  {author} {\bibfnamefont {S.}~\bibnamefont {Dubey}}, \bibinfo {author}
  {\bibfnamefont {Q.}~\bibnamefont {Yao}}, \bibinfo {author} {\bibfnamefont
  {S.}~\bibnamefont {Liu}}, \bibinfo {author} {\bibfnamefont {X.}~\bibnamefont
  {Wang}}, \bibinfo {author} {\bibfnamefont {Q.}~\bibnamefont {Xiong}},
  \bibinfo {author} {\bibfnamefont {L.}~\bibnamefont {Zhang}}, \ and\ \bibinfo
  {author} {\bibfnamefont {A.}~\bibnamefont {Srivastava}},\ }\bibfield  {title}
  {\enquote {\bibinfo {title} {Entanglement of single-photons and chiral
  phonons in atomically thin wse2},}\ }\href {\doibase
  10.1038/s41567-018-0366-7} {\bibfield  {journal} {\bibinfo  {journal} {Nat.
  Phys.}\ }\textbf {\bibinfo {volume} {15}},\ \bibinfo {pages} {221} (\bibinfo
  {year} {2019}{\natexlab{b}})}\BibitemShut {NoStop}%
\bibitem [{\citenamefont {Mella}\ \emph {et~al.}(2023)\citenamefont {Mella},
  \citenamefont {Calvo},\ and\ \citenamefont
  {Foa~Torres}}]{mella2023entangled}%
  \BibitemOpen
  \bibfield  {author} {\bibinfo {author} {\bibfnamefont {J.~D.}\ \bibnamefont
  {Mella}}, \bibinfo {author} {\bibfnamefont {H.~L.}\ \bibnamefont {Calvo}}, \
  and\ \bibinfo {author} {\bibfnamefont {L.~E.}\ \bibnamefont {Foa~Torres}},\
  }\bibfield  {title} {\enquote {\bibinfo {title} {Entangled states induced by
  electron--phonon interaction in two-dimensional materials},}\ }\href
  {\doibase 10.1021/acs.nanolett.3c03316} {\bibfield  {journal} {\bibinfo
  {journal} {Nano Lett.}\ }\textbf {\bibinfo {volume} {23}},\ \bibinfo {pages}
  {11013} (\bibinfo {year} {2023})}\BibitemShut {NoStop}%
\bibitem [{\citenamefont {Lujan}\ \emph {et~al.}(2024)\citenamefont {Lujan},
  \citenamefont {Choe}, \citenamefont {Chaudhary}, \citenamefont {Ye},
  \citenamefont {Nnokwe}, \citenamefont {Rodriguez-Vega}, \citenamefont {He},
  \citenamefont {Gao}, \citenamefont {Nunley}, \citenamefont {Baldini} \emph
  {et~al.}}]{lujan2024spin}%
  \BibitemOpen
  \bibfield  {author} {\bibinfo {author} {\bibfnamefont {D.}~\bibnamefont
  {Lujan}}, \bibinfo {author} {\bibfnamefont {J.}~\bibnamefont {Choe}},
  \bibinfo {author} {\bibfnamefont {S.}~\bibnamefont {Chaudhary}}, \bibinfo
  {author} {\bibfnamefont {G.}~\bibnamefont {Ye}}, \bibinfo {author}
  {\bibfnamefont {C.}~\bibnamefont {Nnokwe}}, \bibinfo {author} {\bibfnamefont
  {M.}~\bibnamefont {Rodriguez-Vega}}, \bibinfo {author} {\bibfnamefont
  {J.}~\bibnamefont {He}}, \bibinfo {author} {\bibfnamefont {F.~Y.}\
  \bibnamefont {Gao}}, \bibinfo {author} {\bibfnamefont {T.~N.}\ \bibnamefont
  {Nunley}}, \bibinfo {author} {\bibfnamefont {E.}~\bibnamefont {Baldini}},
  \emph {et~al.},\ }\bibfield  {title} {\enquote {\bibinfo {title} {Spin--orbit
  exciton--induced phonon chirality in a quantum magnet},}\ }\href {\doibase
  https://doi.org/10.1073/pnas.2304360121} {\bibfield  {journal} {\bibinfo
  {journal} {Proc. Natl. Acad. Sci. U.S.A.}\ }\textbf {\bibinfo {volume}
  {121}},\ \bibinfo {pages} {e2304360121} (\bibinfo {year} {2024})}\BibitemShut
  {NoStop}%
\bibitem [{\citenamefont {Hernandez}\ \emph {et~al.}(2023)\citenamefont
  {Hernandez}, \citenamefont {Baydin}, \citenamefont {Chaudhary}, \citenamefont
  {Tay}, \citenamefont {Katayama}, \citenamefont {Takeda}, \citenamefont
  {Nojiri}, \citenamefont {Okazaki}, \citenamefont {Rappl}, \citenamefont
  {Abramof} \emph {et~al.}}]{hernandez2023observation}%
  \BibitemOpen
  \bibfield  {author} {\bibinfo {author} {\bibfnamefont {F.~G.}\ \bibnamefont
  {Hernandez}}, \bibinfo {author} {\bibfnamefont {A.}~\bibnamefont {Baydin}},
  \bibinfo {author} {\bibfnamefont {S.}~\bibnamefont {Chaudhary}}, \bibinfo
  {author} {\bibfnamefont {F.}~\bibnamefont {Tay}}, \bibinfo {author}
  {\bibfnamefont {I.}~\bibnamefont {Katayama}}, \bibinfo {author}
  {\bibfnamefont {J.}~\bibnamefont {Takeda}}, \bibinfo {author} {\bibfnamefont
  {H.}~\bibnamefont {Nojiri}}, \bibinfo {author} {\bibfnamefont {A.~K.}\
  \bibnamefont {Okazaki}}, \bibinfo {author} {\bibfnamefont {P.~H.}\
  \bibnamefont {Rappl}}, \bibinfo {author} {\bibfnamefont {E.}~\bibnamefont
  {Abramof}},  \emph {et~al.},\ }\bibfield  {title} {\enquote {\bibinfo {title}
  {Observation of interplay between phonon chirality and electronic band
  topology},}\ }\href {\doibase 10.1126/sciadv.adj4074} {\bibfield  {journal}
  {\bibinfo  {journal} {Sci. Adv.}\ }\textbf {\bibinfo {volume} {9}},\ \bibinfo
  {pages} {eadj4074} (\bibinfo {year} {2023})}\BibitemShut {NoStop}%
\bibitem [{\citenamefont {Gao}\ \emph {et~al.}(2023)\citenamefont {Gao},
  \citenamefont {Pan}, \citenamefont {Zhou},\ and\ \citenamefont
  {Zhang}}]{gao2023chiral}%
  \BibitemOpen
  \bibfield  {author} {\bibinfo {author} {\bibfnamefont {Y.}~\bibnamefont
  {Gao}}, \bibinfo {author} {\bibfnamefont {Y.}~\bibnamefont {Pan}}, \bibinfo
  {author} {\bibfnamefont {J.}~\bibnamefont {Zhou}}, \ and\ \bibinfo {author}
  {\bibfnamefont {L.}~\bibnamefont {Zhang}},\ }\bibfield  {title} {\enquote
  {\bibinfo {title} {Chiral phonon mediated high-temperature
  superconductivity},}\ }\href {\doibase 10.1103/PhysRevB.108.064510}
  {\bibfield  {journal} {\bibinfo  {journal} {Phys. Rev. B}\ }\textbf {\bibinfo
  {volume} {108}},\ \bibinfo {pages} {064510} (\bibinfo {year}
  {2023})}\BibitemShut {NoStop}%
\bibitem [{\citenamefont {Baydin}\ \emph {et~al.}(2022)\citenamefont {Baydin},
  \citenamefont {Hernandez}, \citenamefont {Rodriguez-Vega}, \citenamefont
  {Okazaki}, \citenamefont {Tay}, \citenamefont {Noe}, \citenamefont
  {Katayama}, \citenamefont {Takeda}, \citenamefont {Nojiri}, \citenamefont
  {Rappl}, \citenamefont {Abramof}, \citenamefont {Fiete},\ and\ \citenamefont
  {Kono}}]{BaydinPRL2022}%
  \BibitemOpen
  \bibfield  {author} {\bibinfo {author} {\bibfnamefont {A.}~\bibnamefont
  {Baydin}}, \bibinfo {author} {\bibfnamefont {F.~G.~G.}\ \bibnamefont
  {Hernandez}}, \bibinfo {author} {\bibfnamefont {M.}~\bibnamefont
  {Rodriguez-Vega}}, \bibinfo {author} {\bibfnamefont {A.~K.}\ \bibnamefont
  {Okazaki}}, \bibinfo {author} {\bibfnamefont {F.}~\bibnamefont {Tay}},
  \bibinfo {author} {\bibfnamefont {G.~T.}\ \bibnamefont {Noe}}, \bibinfo
  {author} {\bibfnamefont {I.}~\bibnamefont {Katayama}}, \bibinfo {author}
  {\bibfnamefont {J.}~\bibnamefont {Takeda}}, \bibinfo {author} {\bibfnamefont
  {H.}~\bibnamefont {Nojiri}}, \bibinfo {author} {\bibfnamefont {P.~H.~O.}\
  \bibnamefont {Rappl}}, \bibinfo {author} {\bibfnamefont {E.}~\bibnamefont
  {Abramof}}, \bibinfo {author} {\bibfnamefont {G.~A.}\ \bibnamefont {Fiete}},
  \ and\ \bibinfo {author} {\bibfnamefont {J.}~\bibnamefont {Kono}},\
  }\bibfield  {title} {\enquote {\bibinfo {title} {Magnetic control of soft
  chiral phonons in pbte},}\ }\href {\doibase 10.1103/PhysRevLett.128.075901}
  {\bibfield  {journal} {\bibinfo  {journal} {Phys. Rev. Lett.}\ }\textbf
  {\bibinfo {volume} {128}},\ \bibinfo {pages} {075901} (\bibinfo {year}
  {2022})}\BibitemShut {NoStop}%
\bibitem [{\citenamefont {Chumak}\ \emph {et~al.}(2015)\citenamefont {Chumak},
  \citenamefont {Vasyuchka}, \citenamefont {Serga},\ and\ \citenamefont
  {Hillebrands}}]{Chumak2015}%
  \BibitemOpen
  \bibfield  {author} {\bibinfo {author} {\bibfnamefont {A.~V.}\ \bibnamefont
  {Chumak}}, \bibinfo {author} {\bibfnamefont {V.~I.}\ \bibnamefont
  {Vasyuchka}}, \bibinfo {author} {\bibfnamefont {A.~A.}\ \bibnamefont
  {Serga}}, \ and\ \bibinfo {author} {\bibfnamefont {B.}~\bibnamefont
  {Hillebrands}},\ }\bibfield  {title} {\enquote {\bibinfo {title} {Magnon
  spintronics},}\ }\href {\doibase 10.1038/nphys3347} {\bibfield  {journal}
  {\bibinfo  {journal} {Nat. Phys}\ }\textbf {\bibinfo {volume} {11}},\
  \bibinfo {pages} {453–461} (\bibinfo {year} {2015})}\BibitemShut {NoStop}%
\bibitem [{\citenamefont {Demokritov}\ \emph {et~al.}(2006)\citenamefont
  {Demokritov}, \citenamefont {Demidov}, \citenamefont {Dzyapko}, \citenamefont
  {Melkov}, \citenamefont {Serga}, \citenamefont {Hillebrands},\ and\
  \citenamefont {Slavin}}]{Demokritov2006}%
  \BibitemOpen
  \bibfield  {author} {\bibinfo {author} {\bibfnamefont {S.~O.}\ \bibnamefont
  {Demokritov}}, \bibinfo {author} {\bibfnamefont {V.~E.}\ \bibnamefont
  {Demidov}}, \bibinfo {author} {\bibfnamefont {O.}~\bibnamefont {Dzyapko}},
  \bibinfo {author} {\bibfnamefont {G.~A.}\ \bibnamefont {Melkov}}, \bibinfo
  {author} {\bibfnamefont {A.~A.}\ \bibnamefont {Serga}}, \bibinfo {author}
  {\bibfnamefont {B.}~\bibnamefont {Hillebrands}}, \ and\ \bibinfo {author}
  {\bibfnamefont {A.~N.}\ \bibnamefont {Slavin}},\ }\bibfield  {title}
  {\enquote {\bibinfo {title} {Bose–einstein condensation of
  quasi-equilibrium magnons at room temperature under pumping},}\ }\href
  {\doibase 10.1038/nature05117} {\bibfield  {journal} {\bibinfo  {journal}
  {Nature}\ }\textbf {\bibinfo {volume} {443}},\ \bibinfo {pages} {430–433}
  (\bibinfo {year} {2006})}\BibitemShut {NoStop}%
\bibitem [{\citenamefont {Bozhko}\ \emph {et~al.}(2016)\citenamefont {Bozhko},
  \citenamefont {Serga}, \citenamefont {Clausen}, \citenamefont {Vasyuchka},
  \citenamefont {Heussner}, \citenamefont {Melkov}, \citenamefont {Pomyalov},
  \citenamefont {L’vov},\ and\ \citenamefont
  {Hillebrands}}]{bozhko2016supercurrent}%
  \BibitemOpen
  \bibfield  {author} {\bibinfo {author} {\bibfnamefont {D.~A.}\ \bibnamefont
  {Bozhko}}, \bibinfo {author} {\bibfnamefont {A.~A.}\ \bibnamefont {Serga}},
  \bibinfo {author} {\bibfnamefont {P.}~\bibnamefont {Clausen}}, \bibinfo
  {author} {\bibfnamefont {V.~I.}\ \bibnamefont {Vasyuchka}}, \bibinfo {author}
  {\bibfnamefont {F.}~\bibnamefont {Heussner}}, \bibinfo {author}
  {\bibfnamefont {G.~A.}\ \bibnamefont {Melkov}}, \bibinfo {author}
  {\bibfnamefont {A.}~\bibnamefont {Pomyalov}}, \bibinfo {author}
  {\bibfnamefont {V.~S.}\ \bibnamefont {L’vov}}, \ and\ \bibinfo {author}
  {\bibfnamefont {B.}~\bibnamefont {Hillebrands}},\ }\bibfield  {title}
  {\enquote {\bibinfo {title} {Supercurrent in a room-temperature
  bose–einstein magnon condensate},}\ }\href {\doibase 10.1038/nphys3838}
  {\bibfield  {journal} {\bibinfo  {journal} {Nat. Phys.}\ }\textbf {\bibinfo
  {volume} {12}},\ \bibinfo {pages} {1057–1062} (\bibinfo {year}
  {2016})}\BibitemShut {NoStop}%
\bibitem [{\citenamefont {Troncoso}\ and\ \citenamefont
  {N{\'u}nez}(2011)}]{Troncoso2012DynamicsAS}%
  \BibitemOpen
  \bibfield  {author} {\bibinfo {author} {\bibfnamefont {R.~E.}\ \bibnamefont
  {Troncoso}}\ and\ \bibinfo {author} {\bibfnamefont {A.~S.}\ \bibnamefont
  {N{\'u}nez}},\ }\bibfield  {title} {\enquote {\bibinfo {title} {Dynamics and
  spontaneous coherence of magnons in ferromagnetic thin films},}\ }\href
  {\doibase 10.1088/0953-8984/24/3/036006} {\bibfield  {journal} {\bibinfo
  {journal} {J. Phys. Condens. Matter}\ }\textbf {\bibinfo {volume} {24}},\
  \bibinfo {pages} {036006} (\bibinfo {year} {2011})}\BibitemShut {NoStop}%
\bibitem [{\citenamefont {Yuan}\ \emph {et~al.}(2022)\citenamefont {Yuan},
  \citenamefont {Cao}, \citenamefont {Kamra}, \citenamefont {Duine},\ and\
  \citenamefont {Yan}}]{Yuan2022}%
  \BibitemOpen
  \bibfield  {author} {\bibinfo {author} {\bibfnamefont {H.}~\bibnamefont
  {Yuan}}, \bibinfo {author} {\bibfnamefont {Y.}~\bibnamefont {Cao}}, \bibinfo
  {author} {\bibfnamefont {A.}~\bibnamefont {Kamra}}, \bibinfo {author}
  {\bibfnamefont {R.~A.}\ \bibnamefont {Duine}}, \ and\ \bibinfo {author}
  {\bibfnamefont {P.}~\bibnamefont {Yan}},\ }\bibfield  {title} {\enquote
  {\bibinfo {title} {Quantum magnonics: When magnon spintronics meets quantum
  information science},}\ }\href {\doibase 10.1016/j.physrep.2022.03.002}
  {\bibfield  {journal} {\bibinfo  {journal} {Phys. Rep.}\ }\textbf {\bibinfo
  {volume} {965}},\ \bibinfo {pages} {1–74} (\bibinfo {year}
  {2022})}\BibitemShut {NoStop}%
\bibitem [{\citenamefont {Li}\ \emph {et~al.}(2021)\citenamefont {Li},
  \citenamefont {Cao},\ and\ \citenamefont {Yan}}]{Li2021}%
  \BibitemOpen
  \bibfield  {author} {\bibinfo {author} {\bibfnamefont {Z.-X.}\ \bibnamefont
  {Li}}, \bibinfo {author} {\bibfnamefont {Y.}~\bibnamefont {Cao}}, \ and\
  \bibinfo {author} {\bibfnamefont {P.}~\bibnamefont {Yan}},\ }\bibfield
  {title} {\enquote {\bibinfo {title} {Topological insulators and semimetals in
  classical magnetic systems},}\ }\href {\doibase
  10.1016/j.physrep.2021.02.003} {\bibfield  {journal} {\bibinfo  {journal}
  {Physics Reports}\ }\textbf {\bibinfo {volume} {915}},\ \bibinfo {pages}
  {1–64} (\bibinfo {year} {2021})}\BibitemShut {NoStop}%
\bibitem [{\citenamefont {Zhuo}\ \emph {et~al.}(2023)\citenamefont {Zhuo},
  \citenamefont {Kang}, \citenamefont {Manchon},\ and\ \citenamefont
  {Cheng}}]{Zhuo2023}%
  \BibitemOpen
  \bibfield  {author} {\bibinfo {author} {\bibfnamefont {F.}~\bibnamefont
  {Zhuo}}, \bibinfo {author} {\bibfnamefont {J.}~\bibnamefont {Kang}}, \bibinfo
  {author} {\bibfnamefont {A.}~\bibnamefont {Manchon}}, \ and\ \bibinfo
  {author} {\bibfnamefont {Z.}~\bibnamefont {Cheng}},\ }\bibfield  {title}
  {\enquote {\bibinfo {title} {Topological phases in magnonics},}\ }\href
  {http://dx.doi.org/10.1002/apxr.202300054} {\bibfield  {journal} {\bibinfo
  {journal} {Advanced Physics Research}\ } (\bibinfo {year}
  {2023})}\BibitemShut {NoStop}%
\bibitem [{\citenamefont {Huebl}\ \emph {et~al.}(2013)\citenamefont {Huebl},
  \citenamefont {Zollitsch}, \citenamefont {Lotze}, \citenamefont {Hocke},
  \citenamefont {Greifenstein}, \citenamefont {Marx}, \citenamefont {Gross},\
  and\ \citenamefont {Goennenwein}}]{HueblPRL2013}%
  \BibitemOpen
  \bibfield  {author} {\bibinfo {author} {\bibfnamefont {H.}~\bibnamefont
  {Huebl}}, \bibinfo {author} {\bibfnamefont {C.~W.}\ \bibnamefont
  {Zollitsch}}, \bibinfo {author} {\bibfnamefont {J.}~\bibnamefont {Lotze}},
  \bibinfo {author} {\bibfnamefont {F.}~\bibnamefont {Hocke}}, \bibinfo
  {author} {\bibfnamefont {M.}~\bibnamefont {Greifenstein}}, \bibinfo {author}
  {\bibfnamefont {A.}~\bibnamefont {Marx}}, \bibinfo {author} {\bibfnamefont
  {R.}~\bibnamefont {Gross}}, \ and\ \bibinfo {author} {\bibfnamefont
  {S.~T.~B.}\ \bibnamefont {Goennenwein}},\ }\bibfield  {title} {\enquote
  {\bibinfo {title} {High cooperativity in coupled microwave resonator
  ferrimagnetic insulator hybrids},}\ }\href {\doibase
  10.1103/PhysRevLett.111.127003} {\bibfield  {journal} {\bibinfo  {journal}
  {Phys. Rev. Lett.}\ }\textbf {\bibinfo {volume} {111}},\ \bibinfo {pages}
  {127003} (\bibinfo {year} {2013})}\BibitemShut {NoStop}%
\bibitem [{\citenamefont {Tabuchi}\ \emph {et~al.}(2014)\citenamefont
  {Tabuchi}, \citenamefont {Ishino}, \citenamefont {Ishikawa}, \citenamefont
  {Yamazaki}, \citenamefont {Usami},\ and\ \citenamefont
  {Nakamura}}]{TabuchiPRL2014}%
  \BibitemOpen
  \bibfield  {author} {\bibinfo {author} {\bibfnamefont {Y.}~\bibnamefont
  {Tabuchi}}, \bibinfo {author} {\bibfnamefont {S.}~\bibnamefont {Ishino}},
  \bibinfo {author} {\bibfnamefont {T.}~\bibnamefont {Ishikawa}}, \bibinfo
  {author} {\bibfnamefont {R.}~\bibnamefont {Yamazaki}}, \bibinfo {author}
  {\bibfnamefont {K.}~\bibnamefont {Usami}}, \ and\ \bibinfo {author}
  {\bibfnamefont {Y.}~\bibnamefont {Nakamura}},\ }\bibfield  {title} {\enquote
  {\bibinfo {title} {Hybridizing ferromagnetic magnons and microwave photons in
  the quantum limit},}\ }\href {\doibase 10.1103/PhysRevLett.113.083603}
  {\bibfield  {journal} {\bibinfo  {journal} {Phys. Rev. Lett.}\ }\textbf
  {\bibinfo {volume} {113}},\ \bibinfo {pages} {083603} (\bibinfo {year}
  {2014})}\BibitemShut {NoStop}%
\bibitem [{\citenamefont {Ghosh}\ \emph {et~al.}(2023)\citenamefont {Ghosh},
  \citenamefont {Menichetti}, \citenamefont {Katsnelson},\ and\ \citenamefont
  {Polini}}]{GhoshPRB2023}%
  \BibitemOpen
  \bibfield  {author} {\bibinfo {author} {\bibfnamefont {S.}~\bibnamefont
  {Ghosh}}, \bibinfo {author} {\bibfnamefont {G.}~\bibnamefont {Menichetti}},
  \bibinfo {author} {\bibfnamefont {M.~I.}\ \bibnamefont {Katsnelson}}, \ and\
  \bibinfo {author} {\bibfnamefont {M.}~\bibnamefont {Polini}},\ }\bibfield
  {title} {\enquote {\bibinfo {title} {Plasmon-magnon interactions in
  two-dimensional honeycomb magnets},}\ }\href {\doibase
  10.1103/PhysRevB.107.195302} {\bibfield  {journal} {\bibinfo  {journal}
  {Phys. Rev. B}\ }\textbf {\bibinfo {volume} {107}},\ \bibinfo {pages}
  {195302} (\bibinfo {year} {2023})}\BibitemShut {NoStop}%
\bibitem [{\citenamefont {Dyrda\l{}}\ \emph {et~al.}(2023)\citenamefont
  {Dyrda\l{}}, \citenamefont {Qaiumzadeh}, \citenamefont {Brataas},\ and\
  \citenamefont {Barna\ifmmode~\acute{s}\else \'{s}\fi{}}}]{DyrdalPRB2023}%
  \BibitemOpen
  \bibfield  {author} {\bibinfo {author} {\bibfnamefont {A.}~\bibnamefont
  {Dyrda\l{}}}, \bibinfo {author} {\bibfnamefont {A.}~\bibnamefont
  {Qaiumzadeh}}, \bibinfo {author} {\bibfnamefont {A.}~\bibnamefont {Brataas}},
  \ and\ \bibinfo {author} {\bibfnamefont {J.}~\bibnamefont
  {Barna\ifmmode~\acute{s}\else \'{s}\fi{}}},\ }\bibfield  {title} {\enquote
  {\bibinfo {title} {Magnon-plasmon hybridization mediated by spin-orbit
  interaction in magnetic materials},}\ }\href {\doibase
  10.1103/PhysRevB.108.045414} {\bibfield  {journal} {\bibinfo  {journal}
  {Phys. Rev. B}\ }\textbf {\bibinfo {volume} {108}},\ \bibinfo {pages}
  {045414} (\bibinfo {year} {2023})}\BibitemShut {NoStop}%
\bibitem [{\citenamefont {Costa}\ \emph {et~al.}(2023)\citenamefont {Costa},
  \citenamefont {Vasilevskiy}, \citenamefont {Fernández-Rossier},\ and\
  \citenamefont {Peres}}]{Costa2023}%
  \BibitemOpen
  \bibfield  {author} {\bibinfo {author} {\bibfnamefont {A.~T.}\ \bibnamefont
  {Costa}}, \bibinfo {author} {\bibfnamefont {M.~I.}\ \bibnamefont
  {Vasilevskiy}}, \bibinfo {author} {\bibfnamefont {J.}~\bibnamefont
  {Fernández-Rossier}}, \ and\ \bibinfo {author} {\bibfnamefont {N.~M.~R.}\
  \bibnamefont {Peres}},\ }\bibfield  {title} {\enquote {\bibinfo {title}
  {Strongly coupled magnon–plasmon polaritons in graphene-two-dimensional
  ferromagnet heterostructures},}\ }\href {\doibase
  10.1021/acs.nanolett.3c00907} {\bibfield  {journal} {\bibinfo  {journal}
  {Nano Lett.}\ }\textbf {\bibinfo {volume} {23}},\ \bibinfo {pages}
  {4510–4515} (\bibinfo {year} {2023})}\BibitemShut {NoStop}%
\bibitem [{\citenamefont {Bozhko}\ \emph {et~al.}(2020)\citenamefont {Bozhko},
  \citenamefont {Vasyuchka}, \citenamefont {Chumak},\ and\ \citenamefont
  {Serga}}]{Bozhko2020}%
  \BibitemOpen
  \bibfield  {author} {\bibinfo {author} {\bibfnamefont {D.~A.}\ \bibnamefont
  {Bozhko}}, \bibinfo {author} {\bibfnamefont {V.~I.}\ \bibnamefont
  {Vasyuchka}}, \bibinfo {author} {\bibfnamefont {A.~V.}\ \bibnamefont
  {Chumak}}, \ and\ \bibinfo {author} {\bibfnamefont {A.~A.}\ \bibnamefont
  {Serga}},\ }\bibfield  {title} {\enquote {\bibinfo {title} {Magnon-phonon
  interactions in magnon spintronics (review article)},}\ }\href {\doibase
  10.1063/10.0000872} {\bibfield  {journal} {\bibinfo  {journal} {Low Temp.
  Phys.}\ }\textbf {\bibinfo {volume} {46}},\ \bibinfo {pages} {383–399}
  (\bibinfo {year} {2020})}\BibitemShut {NoStop}%
\bibitem [{\citenamefont {Wang}\ \emph
  {et~al.}(2023{\natexlab{a}})\citenamefont {Wang}, \citenamefont {Ren},
  \citenamefont {Hou}, \citenamefont {Cheng},\ and\ \citenamefont
  {Zhang}}]{Wang2023}%
  \BibitemOpen
  \bibfield  {author} {\bibinfo {author} {\bibfnamefont {K.}~\bibnamefont
  {Wang}}, \bibinfo {author} {\bibfnamefont {K.}~\bibnamefont {Ren}}, \bibinfo
  {author} {\bibfnamefont {Y.}~\bibnamefont {Hou}}, \bibinfo {author}
  {\bibfnamefont {Y.}~\bibnamefont {Cheng}}, \ and\ \bibinfo {author}
  {\bibfnamefont {G.}~\bibnamefont {Zhang}},\ }\bibfield  {title} {\enquote
  {\bibinfo {title} {Magnon–phonon coupling: from fundamental physics to
  applications},}\ }\href {\doibase 10.1039/d3cp02683c} {\bibfield  {journal}
  {\bibinfo  {journal} {Phys. Chem. Chem. Phys.}\ }\textbf {\bibinfo {volume}
  {25}},\ \bibinfo {pages} {21802–21815} (\bibinfo {year}
  {2023}{\natexlab{a}})}\BibitemShut {NoStop}%
\bibitem [{\citenamefont {Zheng}\ \emph {et~al.}(2023)\citenamefont {Zheng},
  \citenamefont {Wang}, \citenamefont {Wang}, \citenamefont {Sun},
  \citenamefont {He}, \citenamefont {Yan},\ and\ \citenamefont
  {Yuan}}]{Zheng2023}%
  \BibitemOpen
  \bibfield  {author} {\bibinfo {author} {\bibfnamefont {S.}~\bibnamefont
  {Zheng}}, \bibinfo {author} {\bibfnamefont {Z.}~\bibnamefont {Wang}},
  \bibinfo {author} {\bibfnamefont {Y.}~\bibnamefont {Wang}}, \bibinfo {author}
  {\bibfnamefont {F.}~\bibnamefont {Sun}}, \bibinfo {author} {\bibfnamefont
  {Q.}~\bibnamefont {He}}, \bibinfo {author} {\bibfnamefont {P.}~\bibnamefont
  {Yan}}, \ and\ \bibinfo {author} {\bibfnamefont {H.~Y.}\ \bibnamefont
  {Yuan}},\ }\bibfield  {title} {\enquote {\bibinfo {title} {Tutorial:
  Nonlinear magnonics},}\ }\href {\doibase 10.1063/5.0152543} {\bibfield
  {journal} {\bibinfo  {journal} {J. Appl. Phys.}\ }\textbf {\bibinfo {volume}
  {134}} (\bibinfo {year} {2023}),\ 10.1063/5.0152543}\BibitemShut {NoStop}%
\bibitem [{\citenamefont {Cornelissen}\ \emph {et~al.}(2016)\citenamefont
  {Cornelissen}, \citenamefont {Peters}, \citenamefont {Bauer}, \citenamefont
  {Duine},\ and\ \citenamefont {van Wees}}]{CornelissenPRB2016}%
  \BibitemOpen
  \bibfield  {author} {\bibinfo {author} {\bibfnamefont {L.~J.}\ \bibnamefont
  {Cornelissen}}, \bibinfo {author} {\bibfnamefont {K.~J.~H.}\ \bibnamefont
  {Peters}}, \bibinfo {author} {\bibfnamefont {G.~E.~W.}\ \bibnamefont
  {Bauer}}, \bibinfo {author} {\bibfnamefont {R.~A.}\ \bibnamefont {Duine}}, \
  and\ \bibinfo {author} {\bibfnamefont {B.~J.}\ \bibnamefont {van Wees}},\
  }\bibfield  {title} {\enquote {\bibinfo {title} {Magnon spin transport driven
  by the magnon chemical potential in a magnetic insulator},}\ }\href {\doibase
  10.1103/PhysRevB.94.014412} {\bibfield  {journal} {\bibinfo  {journal} {Phys.
  Rev. B}\ }\textbf {\bibinfo {volume} {94}},\ \bibinfo {pages} {014412}
  (\bibinfo {year} {2016})}\BibitemShut {NoStop}%
\bibitem [{\citenamefont {Nakane}\ and\ \citenamefont
  {Kohno}(2018)}]{NakanePRB2018}%
  \BibitemOpen
  \bibfield  {author} {\bibinfo {author} {\bibfnamefont {J.~J.}\ \bibnamefont
  {Nakane}}\ and\ \bibinfo {author} {\bibfnamefont {H.}~\bibnamefont {Kohno}},\
  }\bibfield  {title} {\enquote {\bibinfo {title} {Angular momentum of phonons
  and its application to single-spin relaxation},}\ }\href {\doibase
  10.1103/PhysRevB.97.174403} {\bibfield  {journal} {\bibinfo  {journal} {Phys.
  Rev. B}\ }\textbf {\bibinfo {volume} {97}},\ \bibinfo {pages} {174403}
  (\bibinfo {year} {2018})}\BibitemShut {NoStop}%
\bibitem [{\citenamefont {Streib}\ \emph {et~al.}(2018)\citenamefont {Streib},
  \citenamefont {Keshtgar},\ and\ \citenamefont {Bauer}}]{StreibPRL2018}%
  \BibitemOpen
  \bibfield  {author} {\bibinfo {author} {\bibfnamefont {S.}~\bibnamefont
  {Streib}}, \bibinfo {author} {\bibfnamefont {H.}~\bibnamefont {Keshtgar}}, \
  and\ \bibinfo {author} {\bibfnamefont {G.~E.~W.}\ \bibnamefont {Bauer}},\
  }\bibfield  {title} {\enquote {\bibinfo {title} {Damping of magnetization
  dynamics by phonon pumping},}\ }\href {\doibase
  10.1103/PhysRevLett.121.027202} {\bibfield  {journal} {\bibinfo  {journal}
  {Phys. Rev. Lett.}\ }\textbf {\bibinfo {volume} {121}},\ \bibinfo {pages}
  {027202} (\bibinfo {year} {2018})}\BibitemShut {NoStop}%
\bibitem [{\citenamefont {Streib}\ \emph {et~al.}(2019)\citenamefont {Streib},
  \citenamefont {Vidal-Silva}, \citenamefont {Shen},\ and\ \citenamefont
  {Bauer}}]{StreibPRB2019}%
  \BibitemOpen
  \bibfield  {author} {\bibinfo {author} {\bibfnamefont {S.}~\bibnamefont
  {Streib}}, \bibinfo {author} {\bibfnamefont {N.}~\bibnamefont {Vidal-Silva}},
  \bibinfo {author} {\bibfnamefont {K.}~\bibnamefont {Shen}}, \ and\ \bibinfo
  {author} {\bibfnamefont {G.~E.~W.}\ \bibnamefont {Bauer}},\ }\bibfield
  {title} {\enquote {\bibinfo {title} {Magnon-phonon interactions in magnetic
  insulators},}\ }\href {\doibase 10.1103/PhysRevB.99.184442} {\bibfield
  {journal} {\bibinfo  {journal} {Phys. Rev. B}\ }\textbf {\bibinfo {volume}
  {99}},\ \bibinfo {pages} {184442} (\bibinfo {year} {2019})}\BibitemShut
  {NoStop}%
\bibitem [{\citenamefont {R\"uckriegel}\ and\ \citenamefont
  {Duine}(2020)}]{RuckriegelPRL2020}%
  \BibitemOpen
  \bibfield  {author} {\bibinfo {author} {\bibfnamefont {A.}~\bibnamefont
  {R\"uckriegel}}\ and\ \bibinfo {author} {\bibfnamefont {R.~A.}\ \bibnamefont
  {Duine}},\ }\bibfield  {title} {\enquote {\bibinfo {title} {Long-range phonon
  spin transport in ferromagnet--nonmagnetic insulator heterostructures},}\
  }\href {\doibase 10.1103/PhysRevLett.124.117201} {\bibfield  {journal}
  {\bibinfo  {journal} {Phys. Rev. Lett.}\ }\textbf {\bibinfo {volume} {124}},\
  \bibinfo {pages} {117201} (\bibinfo {year} {2020})}\BibitemShut {NoStop}%
\bibitem [{\citenamefont {R\"uckriegel}\ \emph {et~al.}(2020)\citenamefont
  {R\"uckriegel}, \citenamefont {Streib}, \citenamefont {Bauer},\ and\
  \citenamefont {Duine}}]{RuckriegelPRB2020}%
  \BibitemOpen
  \bibfield  {author} {\bibinfo {author} {\bibfnamefont {A.}~\bibnamefont
  {R\"uckriegel}}, \bibinfo {author} {\bibfnamefont {S.}~\bibnamefont
  {Streib}}, \bibinfo {author} {\bibfnamefont {G.~E.~W.}\ \bibnamefont
  {Bauer}}, \ and\ \bibinfo {author} {\bibfnamefont {R.~A.}\ \bibnamefont
  {Duine}},\ }\bibfield  {title} {\enquote {\bibinfo {title} {Angular momentum
  conservation and phonon spin in magnetic insulators},}\ }\href {\doibase
  10.1103/PhysRevB.101.104402} {\bibfield  {journal} {\bibinfo  {journal}
  {Phys. Rev. B}\ }\textbf {\bibinfo {volume} {101}},\ \bibinfo {pages}
  {104402} (\bibinfo {year} {2020})}\BibitemShut {NoStop}%
\bibitem [{\citenamefont {Troncoso}\ \emph {et~al.}(2020)\citenamefont
  {Troncoso}, \citenamefont {Bender}, \citenamefont {Brataas},\ and\
  \citenamefont {Duine}}]{TroncosoPRB2020}%
  \BibitemOpen
  \bibfield  {author} {\bibinfo {author} {\bibfnamefont {R.~E.}\ \bibnamefont
  {Troncoso}}, \bibinfo {author} {\bibfnamefont {S.~A.}\ \bibnamefont
  {Bender}}, \bibinfo {author} {\bibfnamefont {A.}~\bibnamefont {Brataas}}, \
  and\ \bibinfo {author} {\bibfnamefont {R.~A.}\ \bibnamefont {Duine}},\
  }\bibfield  {title} {\enquote {\bibinfo {title} {Spin transport in thick
  insulating antiferromagnetic films},}\ }\href {\doibase
  10.1103/PhysRevB.101.054404} {\bibfield  {journal} {\bibinfo  {journal}
  {Phys. Rev. B}\ }\textbf {\bibinfo {volume} {101}},\ \bibinfo {pages}
  {054404} (\bibinfo {year} {2020})}\BibitemShut {NoStop}%
\bibitem [{\citenamefont {Bezuglyj}\ \emph {et~al.}(2019)\citenamefont
  {Bezuglyj}, \citenamefont {Shklovskij}, \citenamefont {Kruglyak},\ and\
  \citenamefont {Vovk}}]{bezuglyj2019spin}%
  \BibitemOpen
  \bibfield  {author} {\bibinfo {author} {\bibfnamefont {A.}~\bibnamefont
  {Bezuglyj}}, \bibinfo {author} {\bibfnamefont {V.}~\bibnamefont
  {Shklovskij}}, \bibinfo {author} {\bibfnamefont {V.}~\bibnamefont
  {Kruglyak}}, \ and\ \bibinfo {author} {\bibfnamefont {R.}~\bibnamefont
  {Vovk}},\ }\bibfield  {title} {\enquote {\bibinfo {title} {Spin seebeck
  effect and phonon energy transfer in heterostructures containing layers of a
  normal metal and a ferromagnetic insulator},}\ }\href {\doibase
  10.1103/PhysRevB.99.134428} {\bibfield  {journal} {\bibinfo  {journal} {Phys.
  Rev. B}\ }\textbf {\bibinfo {volume} {99}},\ \bibinfo {pages} {134428}
  (\bibinfo {year} {2019})}\BibitemShut {NoStop}%
\bibitem [{\citenamefont {Shklovskij}\ \emph {et~al.}(2018)\citenamefont
  {Shklovskij}, \citenamefont {Kruglyak}, \citenamefont {Vovk},\ and\
  \citenamefont {Dobrovolskiy}}]{shklovskij2018role}%
  \BibitemOpen
  \bibfield  {author} {\bibinfo {author} {\bibfnamefont {V.~A.}\ \bibnamefont
  {Shklovskij}}, \bibinfo {author} {\bibfnamefont {V.~V.}\ \bibnamefont
  {Kruglyak}}, \bibinfo {author} {\bibfnamefont {R.~V.}\ \bibnamefont {Vovk}},
  \ and\ \bibinfo {author} {\bibfnamefont {O.~V.}\ \bibnamefont
  {Dobrovolskiy}},\ }\bibfield  {title} {\enquote {\bibinfo {title} {Role of
  magnons and the size effect in heat transport through an insulating
  ferromagnet/insulator interface},}\ }\href {\doibase
  10.1103/PhysRevB.98.224403} {\bibfield  {journal} {\bibinfo  {journal} {Phys.
  Rev. B}\ }\textbf {\bibinfo {volume} {98}},\ \bibinfo {pages} {224403}
  (\bibinfo {year} {2018})}\BibitemShut {NoStop}%
\bibitem [{\citenamefont {An}\ \emph {et~al.}(2013)\citenamefont {An},
  \citenamefont {Vasyuchka}, \citenamefont {Uchida}, \citenamefont {Chumak},
  \citenamefont {Yamaguchi}, \citenamefont {Harii}, \citenamefont {Ohe},
  \citenamefont {Jungfleisch}, \citenamefont {Kajiwara}, \citenamefont {Adachi}
  \emph {et~al.}}]{an2013unidirectional}%
  \BibitemOpen
  \bibfield  {author} {\bibinfo {author} {\bibfnamefont {T.}~\bibnamefont
  {An}}, \bibinfo {author} {\bibfnamefont {V.}~\bibnamefont {Vasyuchka}},
  \bibinfo {author} {\bibfnamefont {K.-i.}\ \bibnamefont {Uchida}}, \bibinfo
  {author} {\bibfnamefont {A.}~\bibnamefont {Chumak}}, \bibinfo {author}
  {\bibfnamefont {K.}~\bibnamefont {Yamaguchi}}, \bibinfo {author}
  {\bibfnamefont {K.}~\bibnamefont {Harii}}, \bibinfo {author} {\bibfnamefont
  {J.}~\bibnamefont {Ohe}}, \bibinfo {author} {\bibfnamefont {M.}~\bibnamefont
  {Jungfleisch}}, \bibinfo {author} {\bibfnamefont {Y.}~\bibnamefont
  {Kajiwara}}, \bibinfo {author} {\bibfnamefont {H.}~\bibnamefont {Adachi}},
  \emph {et~al.},\ }\bibfield  {title} {\enquote {\bibinfo {title}
  {Unidirectional spin-wave heat conveyer},}\ }\href {\doibase
  10.1038/nmat3628} {\bibfield  {journal} {\bibinfo  {journal} {Nat. Mater}\
  }\textbf {\bibinfo {volume} {12}},\ \bibinfo {pages} {549} (\bibinfo {year}
  {2013})}\BibitemShut {NoStop}%
\bibitem [{\citenamefont {Sheikhi}\ \emph {et~al.}(2021)\citenamefont
  {Sheikhi}, \citenamefont {Kargarian},\ and\ \citenamefont
  {Langari}}]{sheikhi2021hybrid}%
  \BibitemOpen
  \bibfield  {author} {\bibinfo {author} {\bibfnamefont {B.}~\bibnamefont
  {Sheikhi}}, \bibinfo {author} {\bibfnamefont {M.}~\bibnamefont {Kargarian}},
  \ and\ \bibinfo {author} {\bibfnamefont {A.}~\bibnamefont {Langari}},\
  }\bibfield  {title} {\enquote {\bibinfo {title} {Hybrid topological
  magnon-phonon modes in ferromagnetic honeycomb and kagome lattices},}\ }\href
  {\doibase 10.1103/PhysRevB.104.045139} {\bibfield  {journal} {\bibinfo
  {journal} {Phys. Rev. B}\ }\textbf {\bibinfo {volume} {104}},\ \bibinfo
  {pages} {045139} (\bibinfo {year} {2021})}\BibitemShut {NoStop}%
\bibitem [{\citenamefont {Park}\ \emph {et~al.}(2020)\citenamefont {Park},
  \citenamefont {Nagaosa},\ and\ \citenamefont {Yang}}]{park2020thermal}%
  \BibitemOpen
  \bibfield  {author} {\bibinfo {author} {\bibfnamefont {S.}~\bibnamefont
  {Park}}, \bibinfo {author} {\bibfnamefont {N.}~\bibnamefont {Nagaosa}}, \
  and\ \bibinfo {author} {\bibfnamefont {B.-J.}\ \bibnamefont {Yang}},\
  }\bibfield  {title} {\enquote {\bibinfo {title} {Thermal hall effect, spin
  nernst effect, and spin density induced by a thermal gradient in collinear
  ferrimagnets from magnon--phonon interaction},}\ }\href {\doibase
  10.1021/acs.nanolett.0c00363} {\bibfield  {journal} {\bibinfo  {journal}
  {Nano Lett.}\ }\textbf {\bibinfo {volume} {20}},\ \bibinfo {pages} {2741}
  (\bibinfo {year} {2020})}\BibitemShut {NoStop}%
\bibitem [{\citenamefont {Zhang}\ \emph {et~al.}(2019)\citenamefont {Zhang},
  \citenamefont {Zhang}, \citenamefont {Okamoto},\ and\ \citenamefont
  {Xiao}}]{zhang2019thermal}%
  \BibitemOpen
  \bibfield  {author} {\bibinfo {author} {\bibfnamefont {X.}~\bibnamefont
  {Zhang}}, \bibinfo {author} {\bibfnamefont {Y.}~\bibnamefont {Zhang}},
  \bibinfo {author} {\bibfnamefont {S.}~\bibnamefont {Okamoto}}, \ and\
  \bibinfo {author} {\bibfnamefont {D.}~\bibnamefont {Xiao}},\ }\bibfield
  {title} {\enquote {\bibinfo {title} {Thermal hall effect induced by
  magnon-phonon interactions},}\ }\href {\doibase
  10.1103/PhysRevLett.123.167202} {\bibfield  {journal} {\bibinfo  {journal}
  {Phys. Rev. Lett.}\ }\textbf {\bibinfo {volume} {123}},\ \bibinfo {pages}
  {167202} (\bibinfo {year} {2019})}\BibitemShut {NoStop}%
\bibitem [{\citenamefont {Li}\ and\ \citenamefont
  {Okamoto}(2022)}]{li2022thermal}%
  \BibitemOpen
  \bibfield  {author} {\bibinfo {author} {\bibfnamefont {S.}~\bibnamefont
  {Li}}\ and\ \bibinfo {author} {\bibfnamefont {S.}~\bibnamefont {Okamoto}},\
  }\bibfield  {title} {\enquote {\bibinfo {title} {Thermal hall effect in the
  kitaev-heisenberg system with spin-phonon coupling},}\ }\href {\doibase
  10.1103/PhysRevB.106.024413} {\bibfield  {journal} {\bibinfo  {journal}
  {Phys. Rev. B}\ }\textbf {\bibinfo {volume} {106}},\ \bibinfo {pages}
  {024413} (\bibinfo {year} {2022})}\BibitemShut {NoStop}%
\bibitem [{\citenamefont {Brinkman}(1967)}]{Brinkman1967}%
  \BibitemOpen
  \bibfield  {author} {\bibinfo {author} {\bibfnamefont {W.}~\bibnamefont
  {Brinkman}},\ }\bibfield  {title} {\enquote {\bibinfo {title} {Magnetic
  symmetry and spin waves},}\ }\href {\doibase 10.1063/1.1709692} {\bibfield
  {journal} {\bibinfo  {journal} {J. Appl. Phys.}\ }\textbf {\bibinfo {volume}
  {38}},\ \bibinfo {pages} {939–943} (\bibinfo {year} {1967})}\BibitemShut
  {NoStop}%
\bibitem [{\citenamefont {Cui}\ \emph {et~al.}(2023)\citenamefont {Cui},
  \citenamefont {Bostr\"{o}m}, \citenamefont {Ozerov}, \citenamefont {Wu},
  \citenamefont {Jiang}, \citenamefont {Chu}, \citenamefont {Li}, \citenamefont
  {Liu}, \citenamefont {Xu}, \citenamefont {Rubio},\ and\ \citenamefont
  {Zhang}}]{Cui2023}%
  \BibitemOpen
  \bibfield  {author} {\bibinfo {author} {\bibfnamefont {J.}~\bibnamefont
  {Cui}}, \bibinfo {author} {\bibfnamefont {E.~V.}\ \bibnamefont
  {Bostr\"{o}m}}, \bibinfo {author} {\bibfnamefont {M.}~\bibnamefont {Ozerov}},
  \bibinfo {author} {\bibfnamefont {F.}~\bibnamefont {Wu}}, \bibinfo {author}
  {\bibfnamefont {Q.}~\bibnamefont {Jiang}}, \bibinfo {author} {\bibfnamefont
  {J.-H.}\ \bibnamefont {Chu}}, \bibinfo {author} {\bibfnamefont
  {C.}~\bibnamefont {Li}}, \bibinfo {author} {\bibfnamefont {F.}~\bibnamefont
  {Liu}}, \bibinfo {author} {\bibfnamefont {X.}~\bibnamefont {Xu}}, \bibinfo
  {author} {\bibfnamefont {A.}~\bibnamefont {Rubio}}, \ and\ \bibinfo {author}
  {\bibfnamefont {Q.}~\bibnamefont {Zhang}},\ }\bibfield  {title} {\enquote
  {\bibinfo {title} {Chirality selective magnon-phonon hybridization and
  magnon-induced chiral phonons in a layered zigzag antiferromagnet},}\ }\href
  {http://dx.doi.org/10.1038/s41467-023-39123-y} {\bibfield  {journal}
  {\bibinfo  {journal} {Nat. Commun.}\ }\textbf {\bibinfo {volume} {14}}
  (\bibinfo {year} {2023})}\BibitemShut {NoStop}%
\bibitem [{\citenamefont {Ma}\ \emph {et~al.}(2023)\citenamefont {Ma},
  \citenamefont {Wang},\ and\ \citenamefont {Chen}}]{ma2023chiral}%
  \BibitemOpen
  \bibfield  {author} {\bibinfo {author} {\bibfnamefont {B.}~\bibnamefont
  {Ma}}, \bibinfo {author} {\bibfnamefont {Z.~D.}\ \bibnamefont {Wang}}, \ and\
  \bibinfo {author} {\bibfnamefont {G.}~\bibnamefont {Chen}},\ }\href@noop {}
  {\enquote {\bibinfo {title} {Chiral magneto-phonons with tunable topology in
  anisotropic quantum magnets},}\ } (\bibinfo {year} {2023}),\ \Eprint
  {http://arxiv.org/abs/2309.04064} {arXiv:2309.04064 [cond-mat.mes-hall]}
  \BibitemShut {NoStop}%
\bibitem [{\citenamefont {Wang}\ \emph
  {et~al.}(2023{\natexlab{b}})\citenamefont {Wang}, \citenamefont {Liu},
  \citenamefont {Long},\ and\ \citenamefont {Wang}}]{WangPRB2023}%
  \BibitemOpen
  \bibfield  {author} {\bibinfo {author} {\bibfnamefont {Q.}~\bibnamefont
  {Wang}}, \bibinfo {author} {\bibfnamefont {S.}~\bibnamefont {Liu}}, \bibinfo
  {author} {\bibfnamefont {M.-Q.}\ \bibnamefont {Long}}, \ and\ \bibinfo
  {author} {\bibfnamefont {Y.-P.}\ \bibnamefont {Wang}},\ }\bibfield  {title}
  {\enquote {\bibinfo {title} {Regulation of magnon-phonon coupling by phonon
  angular momentum in two-dimensional systems},}\ }\href {\doibase
  10.1103/PhysRevB.108.174426} {\bibfield  {journal} {\bibinfo  {journal}
  {Phys. Rev. B}\ }\textbf {\bibinfo {volume} {108}},\ \bibinfo {pages}
  {174426} (\bibinfo {year} {2023}{\natexlab{b}})}\BibitemShut {NoStop}%
\bibitem [{\citenamefont {Yao}\ and\ \citenamefont
  {Murakami}(2024)}]{yao2024conversion}%
  \BibitemOpen
  \bibfield  {author} {\bibinfo {author} {\bibfnamefont {D.}~\bibnamefont
  {Yao}}\ and\ \bibinfo {author} {\bibfnamefont {S.}~\bibnamefont {Murakami}},\
  }\bibfield  {title} {\enquote {\bibinfo {title} {Conversion of chiral phonons
  into magnons in ferromagnets and antiferromagnets},}\ }\href {\doibase
  10.7566/JPSJ.93.034708} {\bibfield  {journal} {\bibinfo  {journal} {J. Phys.
  Soc. Jpn.}\ }\textbf {\bibinfo {volume} {93}},\ \bibinfo {pages} {034708}
  (\bibinfo {year} {2024})}\BibitemShut {NoStop}%
\bibitem [{\citenamefont {Delugas}\ \emph {et~al.}(2023)\citenamefont
  {Delugas}, \citenamefont {Baseggio}, \citenamefont {Timrov}, \citenamefont
  {Baroni},\ and\ \citenamefont {Gorni}}]{delugas2023magnon}%
  \BibitemOpen
  \bibfield  {author} {\bibinfo {author} {\bibfnamefont {P.}~\bibnamefont
  {Delugas}}, \bibinfo {author} {\bibfnamefont {O.}~\bibnamefont {Baseggio}},
  \bibinfo {author} {\bibfnamefont {I.}~\bibnamefont {Timrov}}, \bibinfo
  {author} {\bibfnamefont {S.}~\bibnamefont {Baroni}}, \ and\ \bibinfo {author}
  {\bibfnamefont {T.}~\bibnamefont {Gorni}},\ }\bibfield  {title} {\enquote
  {\bibinfo {title} {Magnon-phonon interactions enhance the gap at the dirac
  point in the spin-wave spectra of cri 3 two-dimensional magnets},}\ }\href
  {\doibase 10.1103/PhysRevB.107.214452} {\bibfield  {journal} {\bibinfo
  {journal} {Phys. Rev. B}\ }\textbf {\bibinfo {volume} {107}},\ \bibinfo
  {pages} {214452} (\bibinfo {year} {2023})}\BibitemShut {NoStop}%
\bibitem [{Note1()}]{Note1}%
  \BibitemOpen
  \bibinfo {note} {The coefficients must be invariant under exchange of lattice
  site indices $i \leftrightarrow j$, so that there is only one tensor of
  coefficients per lattice pair $\{i,j\}$. Also, to be invariant under the
  transformations $\alpha \leftrightarrow \beta $ and $\gamma \leftrightarrow
  \lambda $. These symmetries reduce the number of coefficients to $36$ per
  lattice pair $\{i,j\}$.}\BibitemShut {Stop}%
\bibitem [{\citenamefont {Birss}(1964)}]{birss1964symmetry}%
  \BibitemOpen
  \bibfield  {author} {\bibinfo {author} {\bibfnamefont {R.}~\bibnamefont
  {Birss}},\ }\href {https://books.google.no/books?id=ZjBOxwEACAAJ} {\emph
  {\bibinfo {title} {Symmetry and Magnetism}}},\ \bibinfo {series} {Series of
  monographs on selected topics in solid state physics}\ No.\ \bibinfo {number}
  {v. 3}\ (\bibinfo  {publisher} {North-Holland Publishing Company},\ \bibinfo
  {year} {1964})\BibitemShut {NoStop}%
\bibitem [{\citenamefont {Joshi}(2018)}]{JoshiPRB2018}%
  \BibitemOpen
  \bibfield  {author} {\bibinfo {author} {\bibfnamefont {D.~G.}\ \bibnamefont
  {Joshi}},\ }\bibfield  {title} {\enquote {\bibinfo {title} {Topological
  excitations in the ferromagnetic kitaev-heisenberg model},}\ }\href {\doibase
  10.1103/PhysRevB.98.060405} {\bibfield  {journal} {\bibinfo  {journal} {Phys.
  Rev. B}\ }\textbf {\bibinfo {volume} {98}},\ \bibinfo {pages} {060405}
  (\bibinfo {year} {2018})}\BibitemShut {NoStop}%
\bibitem [{\citenamefont {McClarty}\ \emph {et~al.}(2018)\citenamefont
  {McClarty}, \citenamefont {Dong}, \citenamefont {Gohlke}, \citenamefont
  {Rau}, \citenamefont {Pollmann}, \citenamefont {Moessner},\ and\
  \citenamefont {Penc}}]{McClartyPRB2018}%
  \BibitemOpen
  \bibfield  {author} {\bibinfo {author} {\bibfnamefont {P.~A.}\ \bibnamefont
  {McClarty}}, \bibinfo {author} {\bibfnamefont {X.-Y.}\ \bibnamefont {Dong}},
  \bibinfo {author} {\bibfnamefont {M.}~\bibnamefont {Gohlke}}, \bibinfo
  {author} {\bibfnamefont {J.~G.}\ \bibnamefont {Rau}}, \bibinfo {author}
  {\bibfnamefont {F.}~\bibnamefont {Pollmann}}, \bibinfo {author}
  {\bibfnamefont {R.}~\bibnamefont {Moessner}}, \ and\ \bibinfo {author}
  {\bibfnamefont {K.}~\bibnamefont {Penc}},\ }\bibfield  {title} {\enquote
  {\bibinfo {title} {Topological magnons in kitaev magnets at high fields},}\
  }\href {\doibase 10.1103/PhysRevB.98.060404} {\bibfield  {journal} {\bibinfo
  {journal} {Phys. Rev. B}\ }\textbf {\bibinfo {volume} {98}},\ \bibinfo
  {pages} {060404} (\bibinfo {year} {2018})}\BibitemShut {NoStop}%
\bibitem [{\citenamefont {Holstein}\ and\ \citenamefont
  {Primakoff}(1940)}]{HolsteinPR1940}%
  \BibitemOpen
  \bibfield  {author} {\bibinfo {author} {\bibfnamefont {T.}~\bibnamefont
  {Holstein}}\ and\ \bibinfo {author} {\bibfnamefont {H.}~\bibnamefont
  {Primakoff}},\ }\bibfield  {title} {\enquote {\bibinfo {title} {Field
  dependence of the intrinsic domain magnetization of a ferromagnet},}\ }\href
  {\doibase 10.1103/PhysRev.58.1098} {\bibfield  {journal} {\bibinfo  {journal}
  {Phys. Rev.}\ }\textbf {\bibinfo {volume} {58}},\ \bibinfo {pages} {1098}
  (\bibinfo {year} {1940})}\BibitemShut {NoStop}%
\bibitem [{Note2()}]{Note2}%
  \BibitemOpen
  \bibinfo {note} {It is convenient to choose the quantization axis along the
  saturation magnetization, such that ${\protect \bm {S}}_{\protect \bm
  {r}}={\protect \cal R}_y(\theta )\protect \tilde {\protect \bm {S}}_{\protect
  \bm {r}}$, with ${\protect \cal R}_y(\theta )$ the rotation matrix in an
  angle $\theta $ around $y$ axis $S^x=\protect \tilde {S}^{x'}\cos (\theta
  )+\protect \tilde {S}^{z'}\sin (\theta )$, $S^y=\protect \tilde {S}^{y'}$ and
  $S^z=-\protect \tilde {S}^{x'}\sin (\theta )+\protect \tilde {S}^{z'}\cos
  (\theta )$}\BibitemShut {NoStop}%
\bibitem [{Note3()}]{Note3}%
  \BibitemOpen
  \bibinfo {note} {${\protect \bm {u}}_{\protect \bm {r}}=\DOTSB \sum@
  \slimits@ _{\protect \bm {k},\xi } {\protect \bm {\epsilon }}_{\protect \bm
  {k}\xi }\protect \sqrt {{\hbar }/{2m_{\protect \bm {r}}\omega _{\protect \bm
  {k}\xi }N}}\left (c_{\protect \bm {k}\xi }^\dagger +c_{-\protect \bm {k}\xi
  }\right )e^{i\protect \bm {k}\cdot {\protect \bm {R}}_{\protect \bm
  {r}}}$}\BibitemShut {NoStop}%
\bibitem [{\citenamefont {Colpa}(1978)}]{colpa1978diagonalization}%
  \BibitemOpen
  \bibfield  {author} {\bibinfo {author} {\bibfnamefont {J.}~\bibnamefont
  {Colpa}},\ }\bibfield  {title} {\enquote {\bibinfo {title} {Diagonalization
  of the quadratic boson hamiltonian},}\ }\href {\doibase
  10.1016/0378-4371(78)90160-7} {\bibfield  {journal} {\bibinfo  {journal}
  {Phys. A: Stat. Mech. Appl.}\ }\textbf {\bibinfo {volume} {93}},\ \bibinfo
  {pages} {327} (\bibinfo {year} {1978})}\BibitemShut {NoStop}%
\bibitem [{\citenamefont {Savin}\ and\ \citenamefont
  {Kivshar}(2010)}]{savin2010vibrational}%
  \BibitemOpen
  \bibfield  {author} {\bibinfo {author} {\bibfnamefont {A.~V.}\ \bibnamefont
  {Savin}}\ and\ \bibinfo {author} {\bibfnamefont {Y.~S.}\ \bibnamefont
  {Kivshar}},\ }\bibfield  {title} {\enquote {\bibinfo {title} {Vibrational
  tamm states at the edges of graphene nanoribbons},}\ }\href {\doibase
  10.1103/PhysRevB.81.165418} {\bibfield  {journal} {\bibinfo  {journal} {Phys.
  Rev. B}\ }\textbf {\bibinfo {volume} {81}},\ \bibinfo {pages} {165418}
  (\bibinfo {year} {2010})}\BibitemShut {NoStop}%
\bibitem [{\citenamefont {Simensen}\ \emph {et~al.}(2020)\citenamefont
  {Simensen}, \citenamefont {Kamra}, \citenamefont {Troncoso},\ and\
  \citenamefont {Brataas}}]{broadening2020}%
  \BibitemOpen
  \bibfield  {author} {\bibinfo {author} {\bibfnamefont {H.~T.}\ \bibnamefont
  {Simensen}}, \bibinfo {author} {\bibfnamefont {A.}~\bibnamefont {Kamra}},
  \bibinfo {author} {\bibfnamefont {R.~E.}\ \bibnamefont {Troncoso}}, \ and\
  \bibinfo {author} {\bibfnamefont {A.}~\bibnamefont {Brataas}},\ }\bibfield
  {title} {\enquote {\bibinfo {title} {Magnon decay theory of gilbert damping
  in metallic antiferromagnets},}\ }\href {\doibase
  10.1103/PhysRevB.101.020403} {\bibfield  {journal} {\bibinfo  {journal}
  {Phys. Rev. B}\ }\textbf {\bibinfo {volume} {101}},\ \bibinfo {pages}
  {020403} (\bibinfo {year} {2020})}\BibitemShut {NoStop}%
\end{thebibliography}%

\end{document}